\documentclass[]{interact}

\usepackage{epstopdf}
\usepackage[caption=false]{subfig}

\usepackage[natbibapa,nodoi]{apacite}
\renewcommand\bibliographytypesize{\fontsize{10}{12}\selectfont}

\usepackage{multirow}
\theoremstyle{plain}

\theoremstyle{definition}

\theoremstyle{remark}

\begin{document}

\title{Technical Understanding from IML Hands-on Experience: A Study through a Public Event for Science Museum Visitors}

\author{
\name{Wataru Kawabe\textsuperscript{a}\thanks{CONTACT Wataru Kawabe. Email: wkawabe@iis.u-tokyo.ac.jp}, Yuri Nakao\textsuperscript{b}, Akihisa Shitara\textsuperscript{c}, and Yusuke Sugano\textsuperscript{a}}
\affil{\textsuperscript{a}The University of Tokyo, Tokyo, Japan; 
\textsuperscript{b}Fujitsu Limited, Kawasaki, Japan;
\textsuperscript{c}Tsukuba University, Tsukuba, Japan;
}
}

\maketitle

\begin{abstract}
While AI technology is becoming increasingly prevalent in our daily lives, the comprehension of machine learning (ML) among non-experts remains limited. 
Interactive machine learning (IML) has the potential to serve as a tool for end users, but many existing IML systems are designed for users with a certain level of expertise. 
Consequently, it remains unclear whether IML experiences can enhance the comprehension of ordinary users.
In this study, we conducted a public event using an IML system to assess whether participants could gain technical comprehension through hands-on IML experiences. 
We implemented an interactive sound classification system featuring visualization of internal feature representation and invited visitors at a science museum to freely interact with it. 
By analyzing user behavior and questionnaire responses, we discuss the potential and limitations of IML systems as a tool for promoting technical comprehension among non-experts.
\end{abstract}


\label{sec:introduction}

\section{Introduction}
As artificial intelligence (AI) services and products become increasingly prevalent in our daily lives and their societal impact continues to expand, it is essential for us to understand the underlying technology of AI systems to utilize them effectively.
However, with only a limited number of technical experts involved in AI development, the technology employed in modern AI applications, particularly machine learning (ML) technology, remains a black box to non-expert users who lack specialized knowledge in related fields.
As a result, users often interact solely with the generated outcomes from pre-trained ML models in AI applications~\citep{dudley2018review,kulesza2015principles}. 
Under these circumstances, it becomes challenging for users to develop an interest in or understand ML technology and its technical foundations.

Various interactive machine learning (IML) systems have been developed to make ML technologies accessible and comprehensible for a diverse range of non-expert users~\citep{dudley2018review,fails2003interactive,amershi2014power,ware2001interactive}.
Although these interfaces enable users without advanced technical backgrounds to design ML models, most systems have been developed assuming that users possess some degree of ML knowledge~\citep{mosqueira2022human,pirrung2018sharkzor,wang2018interactive,mosqueira2022human,chromik2021human,goodman2021toward}.
As the technical background is implicitly assumed, these proposed systems have not prioritized engaging users in technical details or promoting comprehension. 
However, since IML is designed to simplify the ML prototyping process, it is also possible that IML systems could facilitate the technical comprehension of end users.
Despite this potential, whether users with limited ML knowledge can familiarize themselves with the technical topics using an IML system has not been fully evaluated.

\begin{figure}[t]
  \centering
  \includegraphics[width=\textwidth]{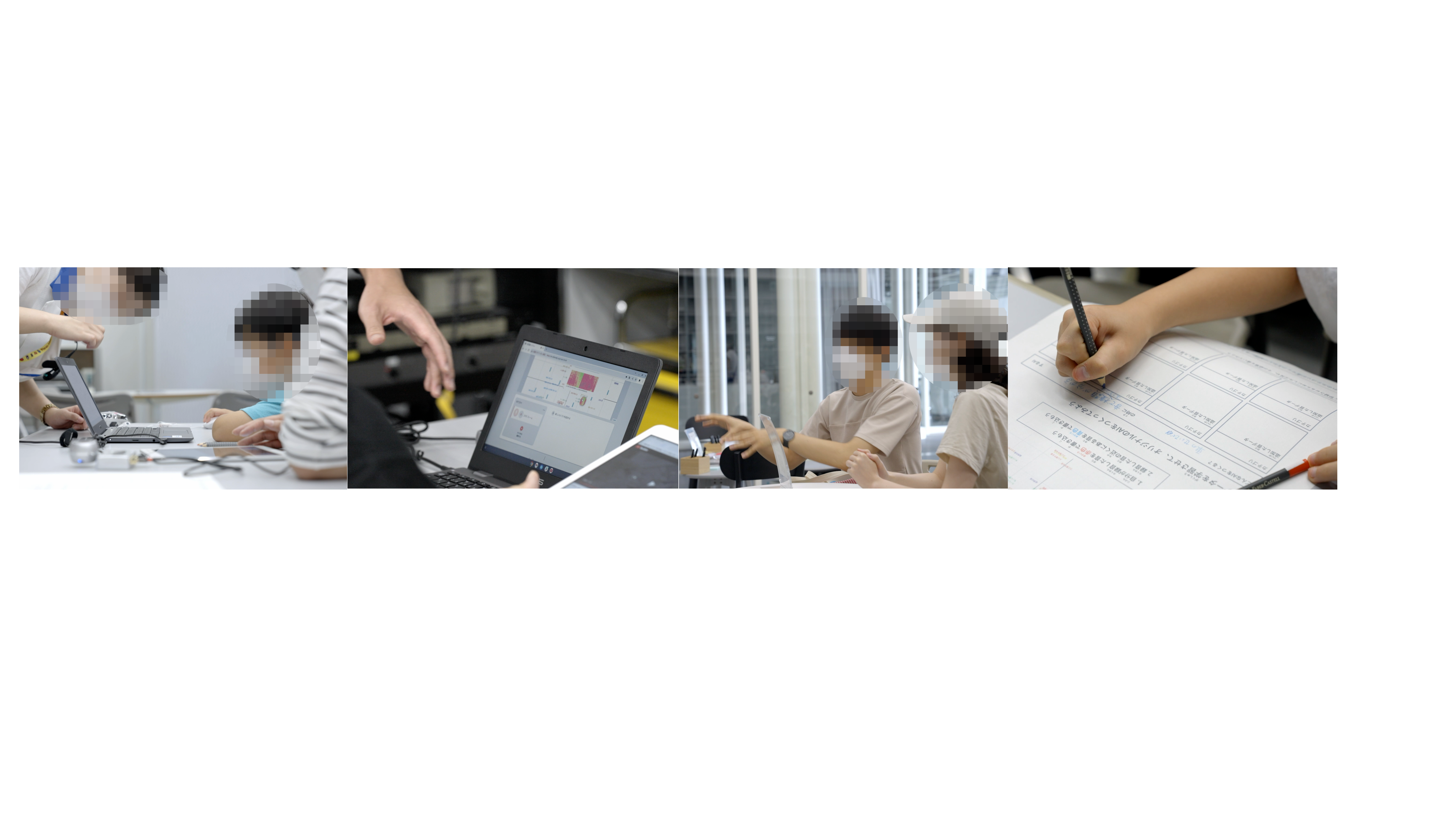}
  \caption{This study investigates the effect of interactive machine learning (IML) on end users' technical comprehension. We hosted a hands-on event in a science museum to invite public visitors.}
  \label{fig:teaser}
\end{figure}

This study aims to accumulate insights into the contributions that IML can make to end users.
As an initial step, we explore how users from the general public acquire technical comprehension, i.e., comprehension of the IML system and ML in general, and thoughts through the operation of an IML system.
To recruit participants from the general public, we chose to host an open event at a science museum and call on visitors to participate (Fig.~\ref{fig:teaser}).
We selected a science museum as the venue because we anticipated that the participants would be more interested in science but not necessarily have specific expertise in AI/ML.
Furthermore, many of the people who experience the event are expected to be families or other groups, and we can also verify the effectiveness of experiencing the IML system while communicating within the group.
Choosing sound classification as the subject, we held an IML hands-on event where the participant groups interacted with a system that supports various explorations and encourages discussion.
We hypothesize that even without detailed instructions, participants could gain insight into the technical points by communicating with each other and using an IML system together.
To test this hypothesis, we allowed them to use the IML system through group discussion freely without imposing quantitative goals or targets.
Using a system that was designed to help the participants understand the ML prototyping process, they experienced interacting with sound data and training a classification model.

We analyzed materials such as interaction logs, post-event questionnaires, comprehension quizzes, and interviews to investigate whether participants became familiar with our system and acquired a basic understanding of ML technology.
Our analysis revealed that participants generally succeeded in grasping the basic concepts of ML and the usage of the system. 
They tended to be confident in understanding these fundamental ideas as well. 
We also identified differences in comprehension between individuals who had previously studied or developed AI and those who had not.
Additionally, the behaviors exhibited during the event demonstrated that participants focused on different technical aspects although sharing the same event.
Finally, while some participants faced misperceptions or insufficient understanding, others reached profound insights that exceeded our expectations.
From these findings, we highlight the potential and limitations of IML as a tool for technical comprehension.

To summarize the contributions, this paper explores the potential of IML systems in fostering technical comprehension among non-expert users. 
By conducting a hands-on event at a science museum, the study examines how participants from the general public can acquire knowledge about ML technology and develop an interest. 
The findings highlight the opportunities and limitations of IML as a tool for technical comprehension.
Additionally, they provide valuable insights for an improved system and event design to engage a broader audience.

\label{sec:related_work}

\section{Related Work}

\subsection{Interactive Machine Learning}

Interactive machine learning (IML)~\citep{dudley2018review,fails2003interactive,amershi2014power,ware2001interactive,kumar2022data} aims to enable users to design personalized ML models through interaction.
IML research aims to enhance task performance efficiently, as IML studies often focus on developing task-specific systems.
For instance, prior research has enabled users to create models for classification~\citep{talbot2009ensemblematrix,carney2020teachable,ishibashi2020investigating,bauerle2022symphony}, data sorting~\citep{pirrung2018sharkzor,hodas2016adding}, visual object recognition~\citep{kacorri2017people,ahmetovic2020recog}, or image segmentation~\citep{wang2018interactive,liao2020iteratively}.
These studies typically conduct user evaluations in controlled environments, where participants are expected to improve the model's accuracy for a specific task.
In contrast, our study primarily focuses on assessing the effectiveness of IML as a tool for understanding the technology rather than solely for efficiently and effectively solving a specific task.

Visualization of large-scale data has been employed in IML studies as a method to facilitate efficient data browsing~\citep{brown2012dis,suh2019anchorviz,kleiman2015dynamicmaps,bian2020deepva,ishibashi2020investigating,chang2021spatial,sivaraman2022emblaze}.
Data visualization is utilized because it has been observed that spatially arranging data points on a two-dimensional surface is conducive to capturing their aggregate trends~\citep{shipman1995finding,malone1983people}.
For instance, organizing images in a two-dimensional space enhances the efficiency of the classification annotator when labeling images compared to examining each image individually~\citep{chang2021spatial}.
Similarly, visualizing and arranging audio samples in a two-dimensional space allows annotators to easily assess the distribution of large-scale sound data~\citep{ishibashi2020investigating}.
We implemented an interactive map that embeds user-recorded sounds alongside pre-registered large-scale samples. 
This study anticipates that such visualization will not only improve annotation efficiency but also aid users in understanding how the system represents sounds.

Several prior studies have also conducted user analyses to understand and improve IML systems~\citep{amershi2009overview,fiebrink2011human,kulesza2009fixing,patel2008investigating,nakao2020use}.
These previous research efforts either tasked users with solving a single pre-defined problem~\citep{kulesza2009fixing,patel2008investigating,fiebrink2011human} or allowed them to design their tasks using the system~\citep{nakao2020use,amershi2009overview}.
The primary focus of these studies was to assess participants' success in creating ML models.
Although these investigations share similarities with our objectives, they typically involved a limited number of participants, ranging from 10 to 20 people~\citep{nakao2020use,amershi2009overview}.
Additionally, as these experiments were conducted in controlled settings, they offered limited insights into users' technical understanding and motivation.
Our study aims to verify the effectiveness of using an IML system with a broader range of participants by conducting public experiments in the more accessible environment of a science museum and inviting visitors on the spot.

\subsection{Technical Understanding for AI/ML}

To help users understand how ML models make inferences, numerous studies have been conducted in the fields of interpretable ML and explainable AI (XAI)~\citep{das2020opportunities,adadi2018peeking,tjoa2020survey,liao2021introduction,sun2022investigating}.
Some of these studies focus on visually representing the model's inner state, enabling users to tune parameters better or create more suitable training data.
For instance, CAM~\citep{zhou2016learning}, Grad-CAM~\citep{selvaraju2017grad}, or Score-CAM~\citep{wang2020score} display the areas that convolutional neural networks focus on for classification as color maps corresponding to input images.
These techniques help users intuitively determine the reasons behind a model's successes or failures.
Other examples include LIME~\citep{ribeiro2016should} and SHAP~\citep{lundberg2017unified}, which identify the input features that strongly contribute to the output.
These approaches retrain interpretable models to approximate the output of the original ML model, providing explanations for specific data points.
Various interpretable ML methods have been attempted to reveal the black-box nature of deep learning~\citep{sundararajan2017axiomatic,ying2019gnnexplainer,barbiero2022entropy}.
While these examples focus on the technical challenges of visualizing model behavior, our study is interested in determining whether the overall IML experience can promote technical understanding.

To foster users' technical understanding, several studies have incorporated XAI technologies within the context of human-computer or human-agent interaction~\cite{silva2022explainable,liao2021human}.
These studies typically focus on domains where enhanced interpretability is beneficial in real-world ML applications, such as medical diagnosis~\cite{nazar2021systematic,oyebode2022machine,wang2019designing}, content design~\cite{eiband2018bringing,zhu2018explainable}, recommendation systems~\cite{herlocker2000explaining,zhang2020explainable}, and decision making~\cite{amini2022discovering}.
Moreover, research has been conducted to determine the most effective methodology for integrating XAI into interactive scenarios~\cite{wolf2019explainability,lim2009assessing,liao2020questioning,xu2023xair}.
These studies focus on how accurately and efficiently users can gain an understanding of AI when provided with explicit explanations about AI interpretability.
Our study, on the contrary, investigates how users acquire understanding through an IML prototyping system.
While our system provides large-scale data visualization and presentation of the model's performance, it does not directly explain its internal mechanism.

There are also studies on the user's technical comprehension of AI, focusing mainly on human factors.
This context mainly includes education for children~\cite{ottenbreit2022principles,wang2022informing,druga2022family,long2022family,dwivedi2021exploring} and opportunities to create personalized applications for non-expert users~\cite{goodman2021toward,kacorri2017people,nakao2020use,kulesza2015principles}.
In the educational context, they investigate the influence of family members on children's learning.
For instance, \cite{druga2022family} studied the parental influence in educating children in AI literacy.
The user-inclusive opportunities encourage users to understand AI/ML through discussions on how to use AI to solve everyday problems.
\cite{goodman2021toward}, for example, investigated how deaf or hard-hearing (DHH) users personalize their assistive sound recognition model to achieve their own goals.
\cite{kulesza2015principles} suggested a framework called Explanatory Debugging, an approach where the system explains its predictions to users, who in turn explain necessary corrections, ultimately enhancing user comprehension.
These studies observe how users interact with ML, given a careful introduction to the system and the event.
Consequently, they have not examined the pure impact of the IML system itself on their comprehension.
In contrast, we provided minimal instructions and refrained from delivering explicit introductions about concepts of AI/ML during the experiment.
By investigating the effects of spontaneous interaction by users, we aim to identify the possibilities and problems of hands-on opportunities using IML systems.

\label{sec:iml_hands-on_event}

\section{IML Hands-on Event}

Our focus lies in determining how hands-on events featuring an IML system can enhance the technical comprehension of non-expert users.
To investigate the effects of unguided exposure to the IML system, we deemed it crucial to allow participants to freely interact with the system based on their objectives rather than prescribing top-down goals.
Consequently, we constructed a natural experimental setup wherein participants received a minimal explanation of the system's capabilities and were permitted to use it autonomously for an unrestricted duration.
Regarding leaving the use of the system up to the user, it is also important to have scenarios where both groups and individuals discuss and use the system.

To conduct such a verification experiment with users from the general public, we chose a science museum and hosted an open hands-on event for visitors there. 
The science museum we selected is visited by people interested in science in general, not just AI/ML, seeking to enjoy various science-related exhibitions.
We aimed to draw these visitors to our event, which was set up inside the exhibition floor, and motivate them to participate.
For the event, we developed an IML system as instructional material, enabling participants to experience the process of training an ML model and interact with the possibly required data.
Following the event, we analyzed participants' interaction logs, questionnaires, quizzes, and post-experience interviews to measure their engagement, comprehension, and thoughts.
In this section, we clarify the details of the system we prepared, the procedure of the event, and the methods to collect the data to be analyzed.

\subsection{Interactive Sound Recognition System}

To offer an opportunity for users to experience IML workflow, we developed a system that facilitates the process of training sound classification models.
It is important for IML workflow that users can experience the process of acquiring training data, and using sound as the input modality can make this process more intuitive.
By realizing the workflow of playing a video on the Internet and recording its sound into the system, the user can work without going through the abstract process of uploading and downloading.
One key aspect of our system is a map representing the distribution of sound features.
Embedding-based feature maps are useful for efficiently grasping the attributes of large datasets~\citep{ishibashi2020investigating,sivaraman2022emblaze}. 
Our system also employs this technique as a tool to help users better understand how the sound is represented within the system.

\subsubsection{UI Overview and Interaction Flow}

\begin{figure}[t]
    \centering
    \includegraphics[width=0.8\linewidth]{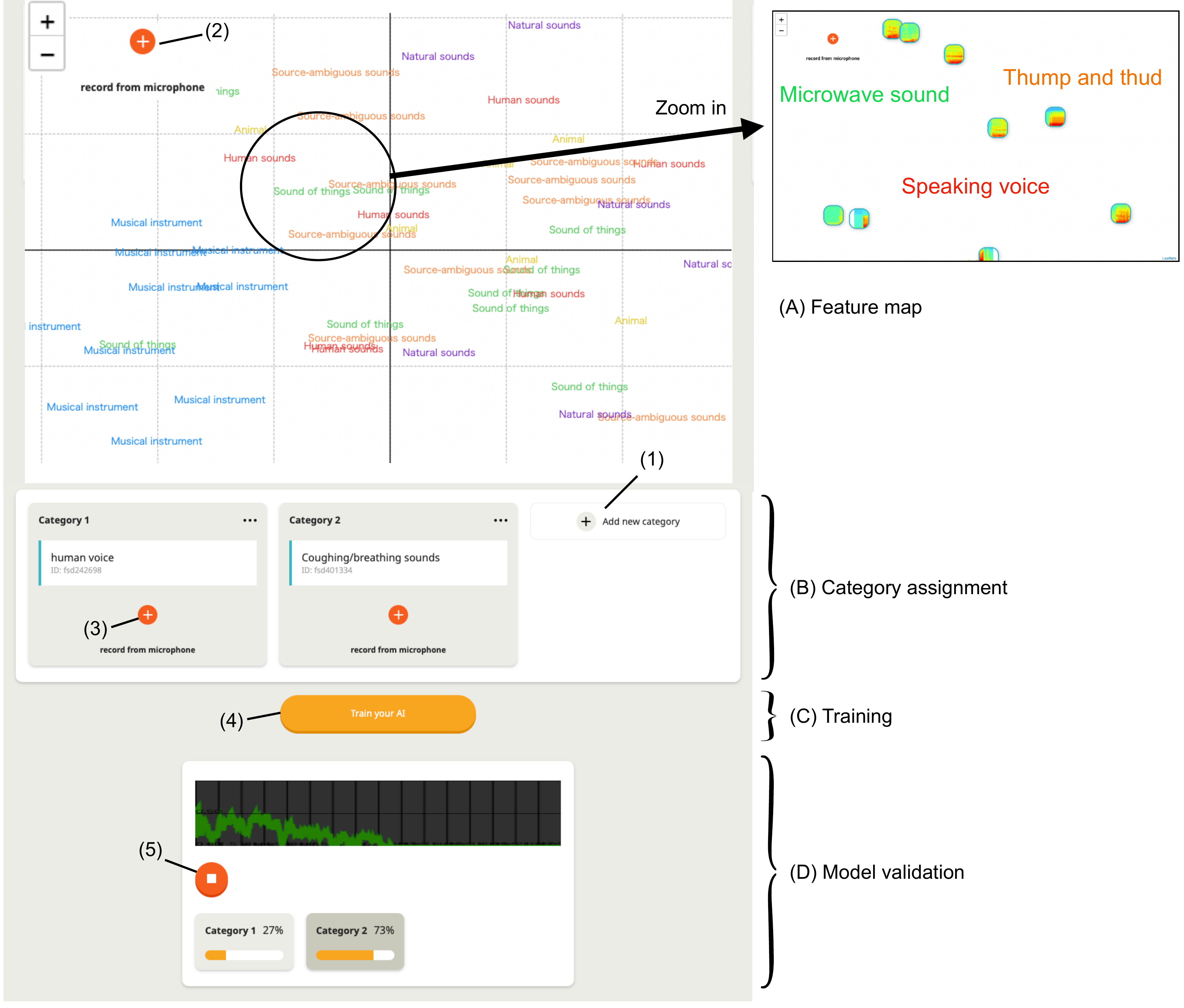}
    \caption{The overview of our IML system (translated from Japanese). The system has three major components, allowing users to create sound classification models.}
    \label{fig:ui_overview}
\end{figure}

Figure~\ref{fig:ui_overview} shows the UI overview of our IML system.
The user interface was implemented using Japanese, the native language of the country where the event was held.
The system consists of four major components: (A) feature map, (B) category assignment, (C) training, and (D) model validation.
The system's ultimate goal is to allow users to create their original sound classification models.
Hence, central to this process is the category assignment component, which allows users to set up their classification categories and assign any sample to each category.
Users can add their classification categories in the category assignment area by clicking the button (1).
Then users can either choose features from the feature map or record new sounds by clicking buttons (2) or (3), respectively.
When users click the button (4), the model is trained with the user-defined sound-category pairs.
Users can also check the real-time inference results by clicking the button (5) in the model validation area.

As is common in training general-purpose recognition models, our ML model uses a pre-trained sound feature extractor on the backend.
One of the purposes of the map component is to visualize such internal mechanisms of IML systems.
By default, the map displays sound samples taken from the FSD50K dataset~\citep{fonseca2021fsd50k}.
We randomly extract $5,000$ public domain data from the FSD50K and pre-compute their sound features and two-dimensional t-SNE embeddings~\citep{van2008visualizing}.
The feature map plots the samples according to the embedding coordinates, and users can move the display area of the map with mouse operation.
The samples are hierarchically clustered and displayed adaptively according to the user's zoom level.
Following a previous study~\citep{ishibashi2020investigating}, individual sounds displayed at the maximum zoom level are visualized using spectrograms.
When the user clicks individual spectrograms, the pop-up window shows a sound playback component and the original category names defined in the FSD50K dataset. 
When zoomed out, the map displays sound clusters.
In this case, pop-up windows display representative categories of the clusters and playback components for a few representative sounds.
In addition, to allow users to catch a glimpse of the samples in the display area, the category names of samples present in the vicinity are displayed in the background.
User-recorded sounds are also displayed on the same map using the algorithm to embed novel samples to the reference t-SNE space~\citep{polivcar2021embedding}.

\begin{figure}[t]
    \centering
    \includegraphics[width=0.95\linewidth]{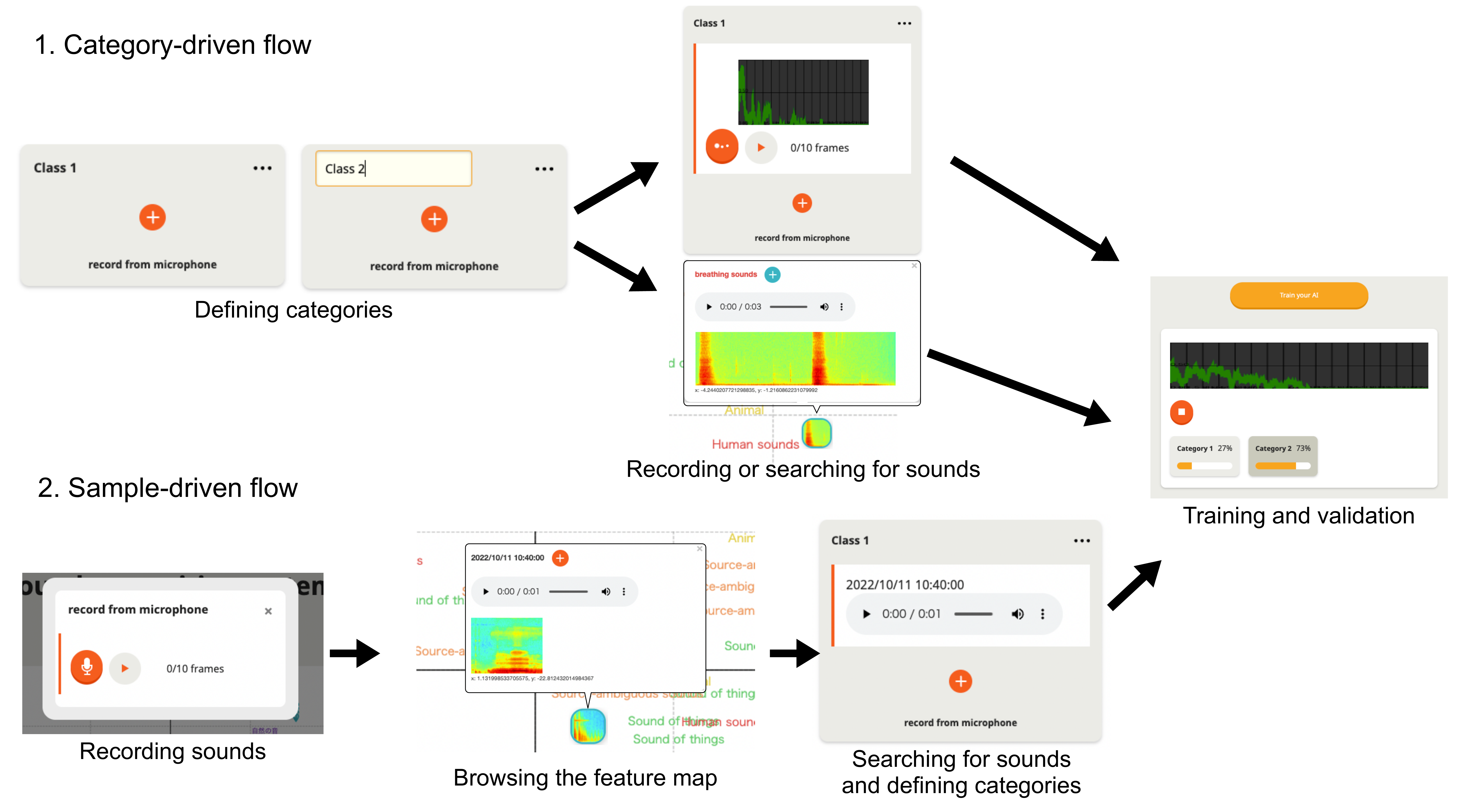}
    \caption{The interaction flow of our system (translated). In the category-driven flow, users first define target categories and then search for appropriate sound samples on the map or directly record the sound. In the sample-driven flow, users first record a sound on the feature map and then register sounds to the categories. They then train the classification model with the training data and validate its performance in both cases.}
    \label{fig:interaction_flow}
\end{figure}

Figure~\ref{fig:interaction_flow} shows the interaction flow of the system.
Our system assumes two patterns as the flow of recording and category assignment and is implemented so that both patterns are possible in parallel.
One is a category-driven top-down flow that begins with the definition of the category by the user.
In this flow pattern, it is assumed that the user has a clear picture of the classification model they want to create and can define the categories at the outset.
The user can either record new sounds directly under the defined categories (button (3) in Fig.~\ref{fig:ui_overview}) or assign categories by searching for sounds that match their purpose on the map.
The sound data recorded this way is displayed on the map with its category name in the pop-up window.

The other is a sample-driven bottom-up flow where the user first records new sounds directly on the map.
For non-expert users, coming up with a classification task is not always an easy task.
Based on this observation, we implement our system in a way that allows users to record sounds without specifying a category. 
By pressing the button in the upper left corner of the map (button (2) in Fig.~\ref{fig:ui_overview}), the users can record sounds as they like.
The pop-up window displays the recording timestamp for such sound data.
This operation allows them to think about classification categories and experience how the recorded sound is mapped to the feature space itself. 
We expect that these opportunities also have a learning effect on users.

\subsubsection{System Implementation}

Our IML system comprised a frontend based on Node.js, a Python/Flask-based ML backend, and a PostgreSQL database.
The above user interface was implemented in JavaScript as a web application.
The system was deployed on a cloud server, and each PC at the event accessed the front-end server from a browser.

The classification model was implemented on the backend server using TensorFlow~\citep{abadi2016tensorflow} and scikit-learn~\citep{pedregosa2011scikit} libraries.
We used the VGGish feature extractor pre-trained on the AudioSet dataset~\citep{hershey2017cnn,gemmeke2017audio}.
The implementation of the feature extraction process follows the official TensorFlow sample code and the prior study~\citep{nakao2020use}.
Briefly, the sound was first converted into a mel-spectrogram, and CNN features were calculated with a one-second sliding window.
To achieve fast training, user-defined classification models were implemented to be trained with the Random Forest algorithm~\citep{breiman2001random}.
The input was the 128-dimensional feature extracted from the feature extractor, and the Random Forest used the default parameters of the scikit-learn library~\citep{pedregosa2011scikit} except that the number of trees was set to 15.

The map embedding was implemented using the openTSNE library~\citep{polivcar2019opentsne}.
The initial map was pre-computed by embedding 128-dimensional feature vectors of the random $5,000$ FSD50K samples into a two-dimensional space.
The feature vector representing each sound data was computed by taking the temporal average of the original features.
The initial embeddings were first clustered into 500 clusters using the $k$-means algorithm~\citep{lloyd1982least}, and their centroid locations were sequentially clustered into 100 and 20 clusters.
The category aggregation results for the included sounds and the top 20 sound IDs closest to the cluster centroid were stored as information for each cluster.
The frontend UI displays playback components for the top three representative sounds.
The same 128-dimensional average features were calculated from the newly recorded sounds, and their coordinates were computed using the Poli{\v{c}}ar et al.'s algorithm~\citep{polivcar2021embedding} implemented in the openTSNE library.

\subsection{Procedure}

The event was held for two days in a room on one of the exhibition floors at a science museum.
The room had a glass wall, and museum visitors could freely see what was happening inside.
We posted an announcement on the museum website and social network, but no pre-registration was requested to attend the event.
A sign about the event was also placed at the room's entrance so that museum visitors could participate in the event spontaneously.
Data collection for the experiments was conducted under the approval of the university's ethics review committee and the science museum.

We further provided the system's demonstration at the event room's entrance.
One of our staff members operated the system on a large screen and created a model that classifies the sounds made by miscellaneous items at hand, such as bells and chimes.
The demonstration also included a brief introduction to the feature map.
The staff provided basic explanations, including that the sounds are mapped according to their characteristics and that sounds with similar features are displayed closer together.

If a group of museum visitors became interested in the event after the demonstration, they consented to data collection and moved on to using the system.
Most of the participants visited the museum in groups, and the following process was conducted with one group as a unit.
Group participants were represented by one person who operated the system and answered the questionnaire, while the others were not directly involved in the work.
Throughout the work, we allowed discussion within the group at any time since our interest was to evaluate their total experience, including an active discussion among members.

We prepared five laptops in the event room, with one laptop per group of participants.
Each laptop was connected to a mouse, earphones for previewing sounds, and an audio interface.
Although our system is designed to record sound from a microphone, the recording will contain a large environmental noise when multiple groups make sound simultaneously in the same room.
There is also a significant limitation to the types of sounds that can be recorded indoors.
Therefore, we connected a tablet PC to the audio interface instead of a microphone so that participants can search and play online videos to record diverse sounds.
The system's operation was left to the participants, but staff members were also available to answer if they had a question.

\begin{figure}[t]
    \centering
    \subfloat[The first half of the worksheet shows the \\feature map and instructs the participant to write \\down sound locations.]{
    \resizebox*{7cm}{!}{\includegraphics{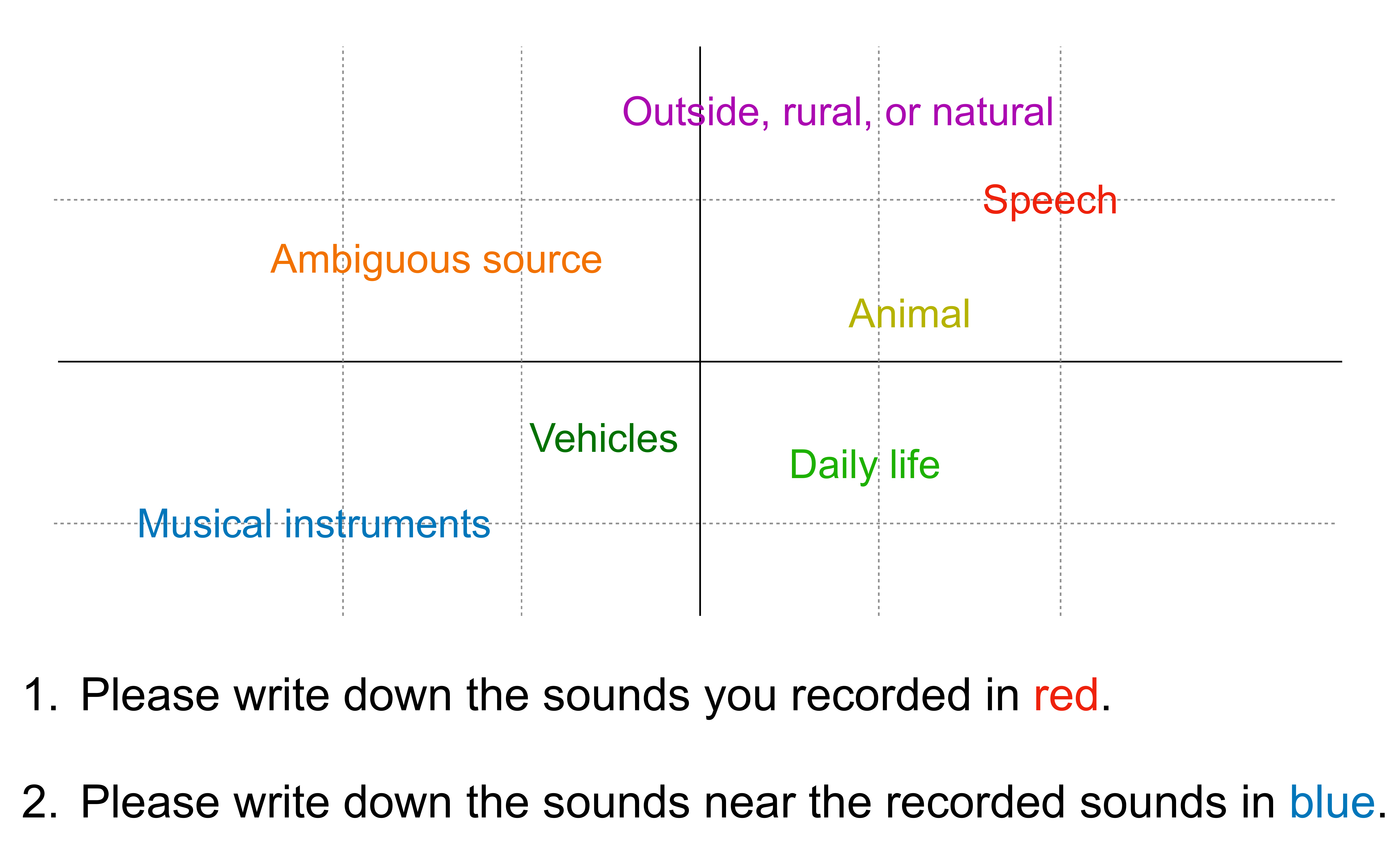}}}
    \subfloat[The second half of the worksheet guides participants to think about their own sound classification models.]{
    \resizebox*{7cm}{!}{\includegraphics{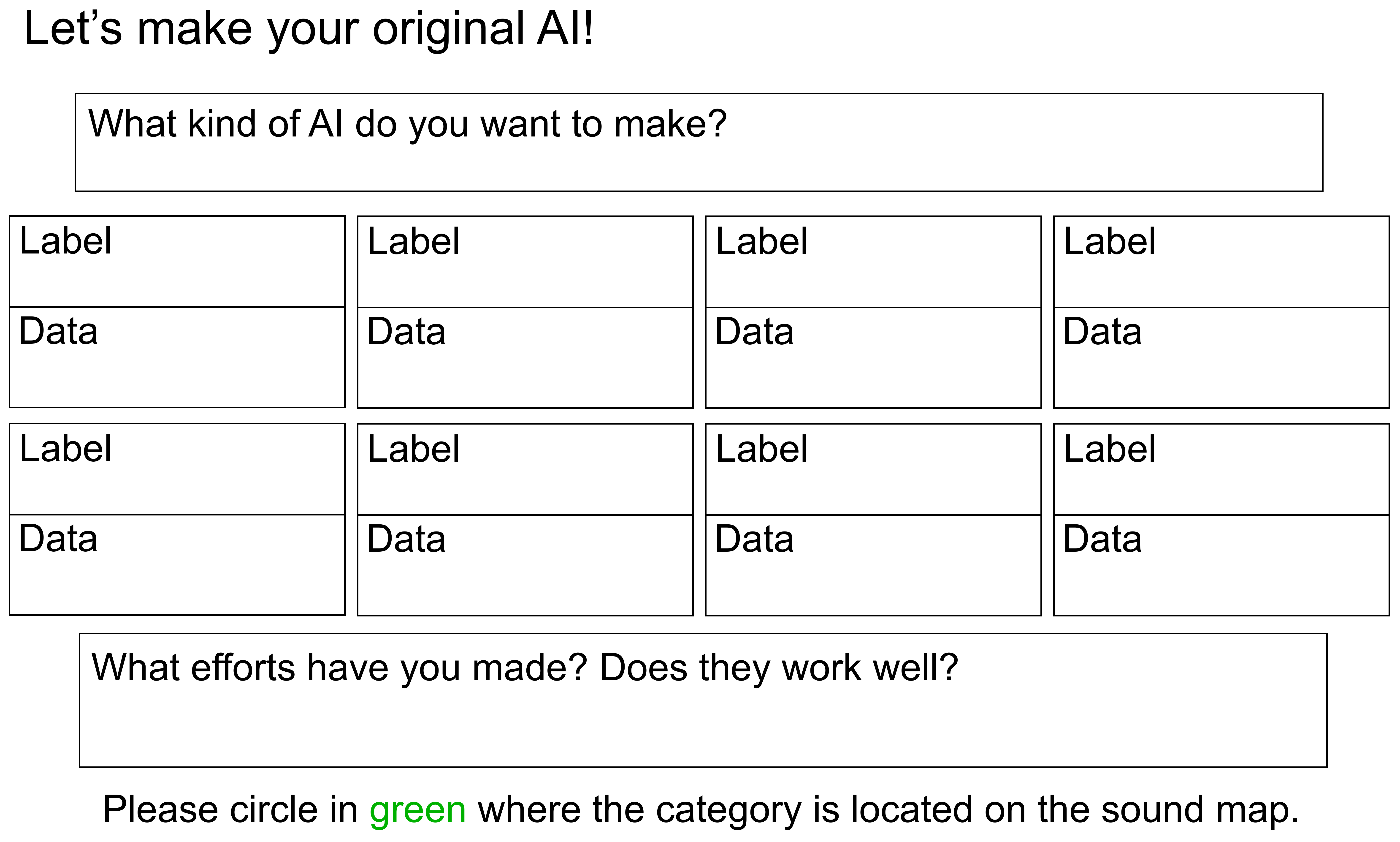}}}
    \caption{Illustrations of the worksheet used at the event (translated from Japanese).} 
    \label{fig:worksheet}
\end{figure}

We also provided worksheets to assist participants in using the system and to allow them to summarize their findings and thoughts.
Figure~\ref{fig:worksheet} shows the worksheet overview containing two sections.
The first half (Fig.~\ref{fig:worksheet}(a)) shows the same feature map as the system and instructs the participant to write down where their recorded sounds were mapped and what sounds were around them.
The second half (Fig.~\ref{fig:worksheet}(b)) is intended to serve as a template for participants to think about sound classification models.
Participants were expected to briefly describe what kind of AI models they wanted to create.
Then they described specific classification categories below and what kind of training data they have added to each category.
They could also take notes during their trial at the bottom, and the worksheet also instructs participants to mark where the category they were trying to classify exists on the map.
Although we encouraged participants to fill out the worksheet, we made it optional.

\subsection{Data Collection}

We evaluated participants' understanding and thoughts using a multifaceted approach, which included logging their interactions and having them complete questionnaires and quizzes. 
In-depth interviews were also conducted with consenting participants. 
Participants were allowed to collaborate on responses for groups such as families or couples, as our goal was to assess the overall experience, including group discussions.
We provide details on the format of each assessment component and how they were presented to the participants.

\subsubsection*{Interaction Logs}

To grasp the user activities of event participants, we recorded their interaction logs while they were using the system.
Timestamped logs were recorded when the participant triggered the following interactions: start/stop recording a sound, delete a recorded sound, open a popup window of a sound, play the audio sample from the popup, train the model, and start/stop the model validation.

\subsubsection*{Usability Questions}

After the session, we asked participants to answer usability questionnaires.
The objective was to evaluate whether the participants could experience the system in the way we intended.
We prepared the questionnaires as an online form, and the participants were asked to answer each item on a tablet provided by us or on their smartphones.
We presented the following questions and asked representative persons from each group to provide five-point Likert scale ratings from \emph{strongly disagree} to \emph{strongly agree}.

\begin{description}
    \item[UQ1] You could understand what the feature map means.
    \item[UQ2] You could use the sounds originally arranged in the feature map to train the AI model.
    \item[UQ3] You could easily find the sounds you had recorded in the feature map.
    \item[UQ4] You could understand how an AI model distinguishes sounds.
    \item[UQ5] You could create an AI model to make decisions the way you wanted.
\end{description}

\subsubsection*{Comprehension Quizzes}

In the same form, we also asked comprehension quizzes to evaluate whether the participants could gain technical understanding after the event.
We prepared simple quizzes to determine whether participants have reached a rudimentary understanding of the system and ML.
To flexibly measure their understanding, we let participants answer in an open-ended format.

\begin{description}
    \item[CQ1] What do you need to make the AI model learn sounds?
    \item[CQ2] How is the position of the recorded sound on the feature map determined?
    \item[CQ3] What is a good practice to improve the performance of the AI model?
\end{description}

The quizzes are related to the general knowledge of training an ML model (CQ1 and CQ3) and embedding data (CQ2), the latter of which is also an indispensable factor in understanding large-scale data-driven ML methods.
Although we originally prepared four quizzes, one of them was excluded because it did not convey our intent and yielded several answers that could not be undoubtedly determined to be correct.
We also asked participants to rate their confidence for each quiz on a 5-point Likert scale.
This is because the open-ended answers alone cannot guarantee their certainty, and we assume that confidence rating is also meaningful in assessing their technical understanding.

We prepared sample answers in advance for each question.
For CQ1, sample answers were either about \textit{recording sounds, making an AI model memorize sounds} or \textit{updating parameters of an AI model}.
The sample answers for CQ2 were also twofold: \textit{referring to the similarities between sounds} or \textit{referring to the particular aspect such as frequency or loudness}.
CQ3 should be related to designing training data, and sample answers were \textit{increasing the amount of the variety of sounds} or \textit{recording better quality sounds}.
A participant's answer was considered correct if it fell into one of these sample answers.

\subsubsection*{Interviews}

After the event, we conducted brief semi-structured interviews with participants who agreed.
The objective was to analyze their technical understanding and system usage further.
The question items were the same as the ones in the quizzes, but we asked further questions in response to their statements.
We interviewed eight participant groups, and one representative person from each group answered the interview while discussing questions with other group members.

\label{sec:results}

\section{Results}

\begin{table}
  \tbl{Summary of the participant demographics.}  
  {\begin{tabular}{lccccc}
    \toprule
    AI-related Experience & Male & Female & Other & N/A & Total \\
    \midrule 
     None & 21 & 20 & 0 & 0 & 41 \\ 
    Studying AI & 10 & 5 & 1 & 1 & 17 \\
    Developing AI & 0 & 0 & 1 & 0 & 1 \\
    Both & 2 & 0 & 0 & 0 & 2 \\
    \midrule
    Total & 33 & 25 & 2 & 1 & 61 \\
    \bottomrule
    \end{tabular}}
    \label{tab:participants}
\end{table}

In the questionnaire, we collected demographic information from participants, including gender, previous experience studying AI systems or models, and previous experience developing AI systems or models.
Table~\ref{tab:participants} represents the demographies of event participants who completed the questionnaire.
Representatives from each group, 63 in total, answered the questionnaire, but two of them were excluded from the analyses because the interaction logs were not saved appropriately.
The representative persons (33 male, 25 female, 2 other, 1 N/A) ranged from 7 to 49 years old ($M = 17.37$, $SD = 9.74$).
41 of them had never learned or implemented anything related to AI or ML.
Out of the 20 participants who had experienced AI before, 17 reported having studied the basic concept without any development experience. 
One participant had only development experience, while the other had both study and development experiences.

For Likert scale questions, we converted the answers to numerical values and calculated mean scores to highlight differences among the items.
In this conversion, \textit{strongly agree} was converted to 2 points, \textit{agree} to 1 point, \textit{neutral} to 0 points, \textit{disagree} to -1 point, and \textit{strongly disagree} to -2 points, respectively.
In the following part, ``score'' means the mean scores for each question from each group of participants.
For the comprehension quiz, we determined the percentage of correct answers for each item, as participants' responses could only be classified as correct or incorrect.

\subsection{Overall Trends}

\begin{figure}[t]
    \centering
    \subfloat[Scores for the usability questions.]{
    \resizebox*{7cm}{!}{\includegraphics{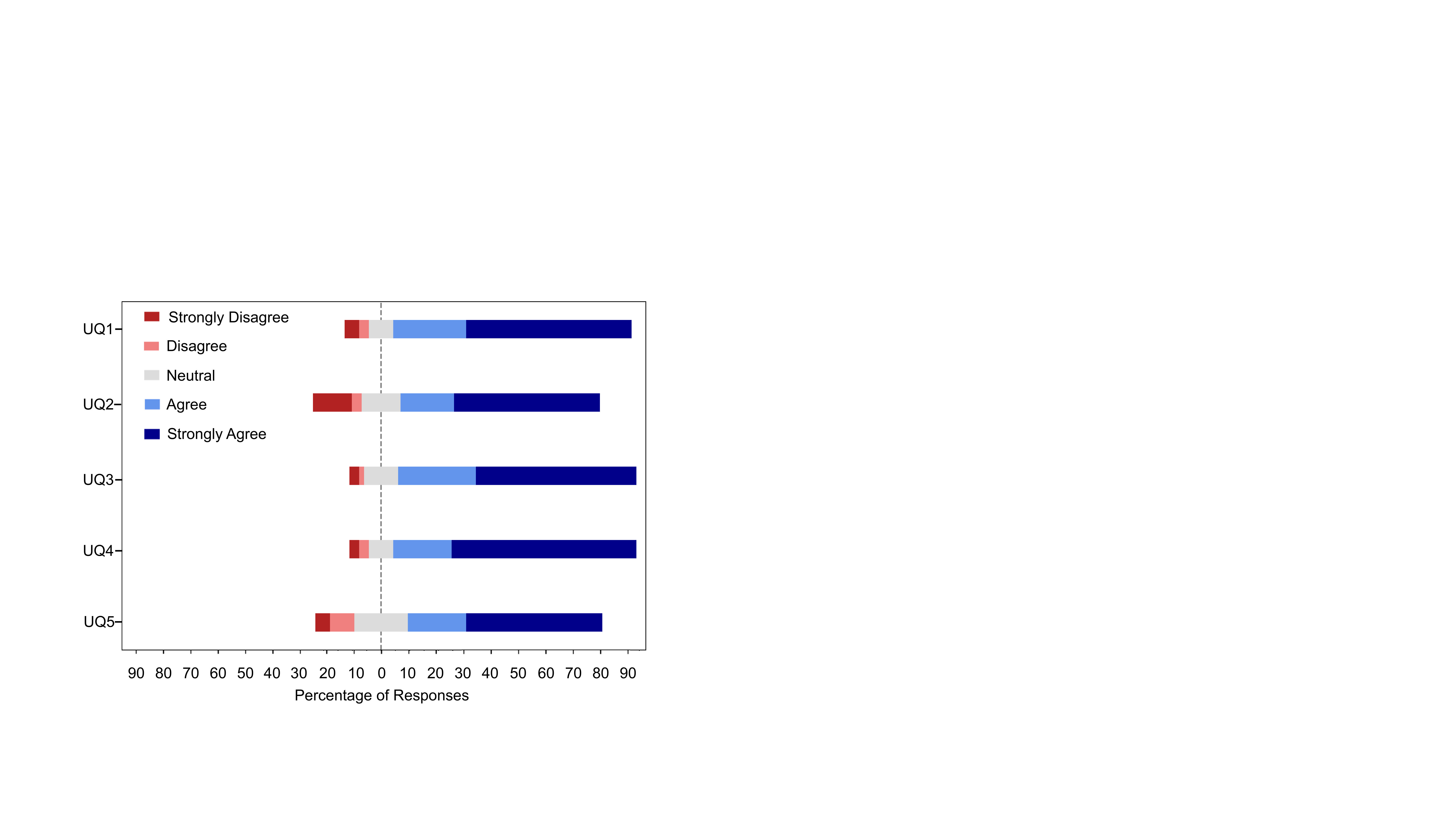}}}
    \subfloat[Percentage of correct answers (above) and confidence scores for the comprehension quizzes (below).]{
    \resizebox*{7cm}{!}{\includegraphics{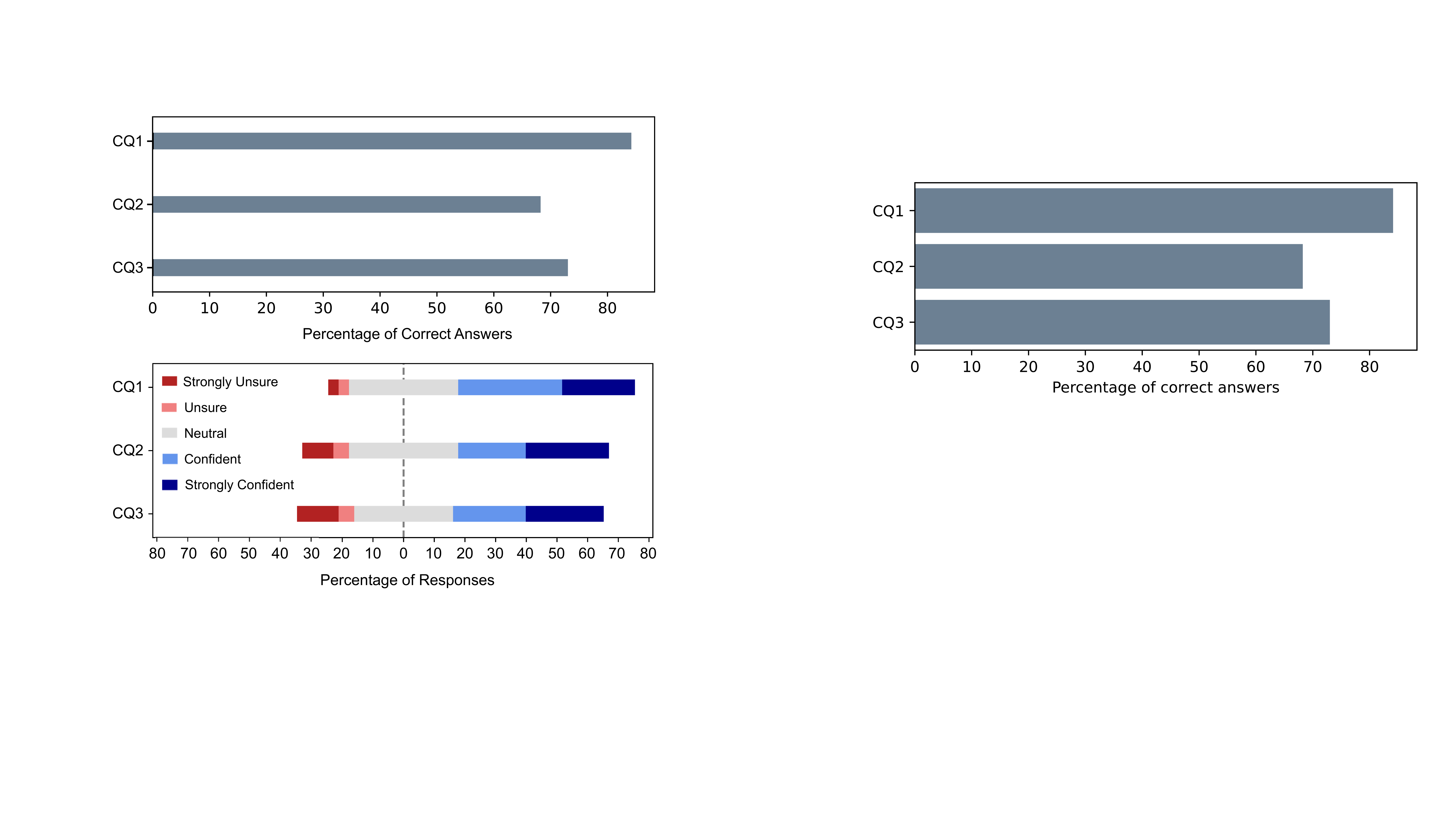}}}
    \caption{Overall trends in questionnaire responses.} 
    \label{fig:general_trends}
\end{figure}

Figure~\ref{fig:general_trends}(a) displays the result of the usability questions.
Overall, participants rated the system highly in terms of usability.
When comparing individual items, UQ2, which focused on whether participants could use the samples on the map as training data, received a relatively lower score ($0.86$) compared to other questions ($1.24, 1.25, 1.36, 0.96$ for UQ1, UQ3, UQ4, and UQ5, respectively).
The difference in operating patterns for obtaining training data, either picking up samples from the map or directly recording, suggests that some participants relied exclusively on recordings, while others made use of the map for sound collection.

The upper portion of Fig.~\ref{fig:general_trends}(b) shows the percentage of correct answers for each quiz item, ($84\%$, $68\%$, and $73\%$, for CQ1, 2, and 3).
The scores suggest that CQ1 (how to train the model) was more straightforward than CQ2 (how the embeddings are decided) and CQ3 (how to improve the accuracy).
One possible explanation is that information related to CQ1 was presented more explicitly during the event compared to the other quiz items.
In contrast, despite the absence of explicit suggestions for the remaining quizzes during the event, many participants were still able to arrive at the correct answers.
Note that the coding results for the quiz are presented in the \textit{total} column of Tab.~\ref{tab:coding}.
The correct answers to each item can be roughly divided into two ways of answering.
The lower half of Fig.~\ref{fig:general_trends}(b) illustrates the overview of confidence ratings.
Overall, participants' confidence levels showed similar trends to the average percentage of correct answers. 
CQ1 exhibited higher mean confidence ($0.72$) than CQ2 ($0.51$) and CQ3 ($0.42$).

\subsection{Differences based on Prior Knowledge}

\begin{figure}[t]
    \centering
    \subfloat[Scores for the usability questions.]{
    \resizebox*{7cm}{!}{\includegraphics{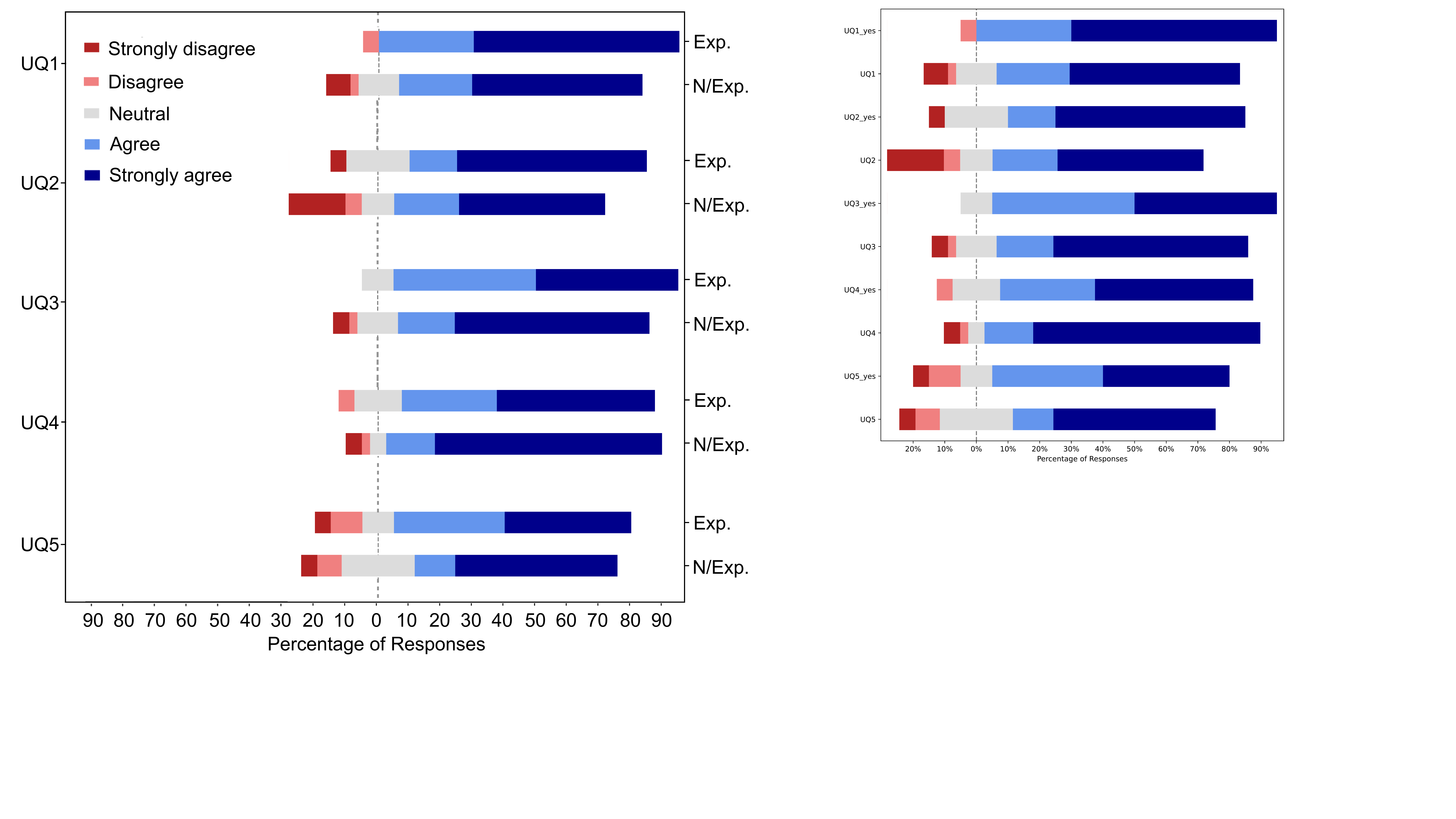}}}
    \subfloat[Percentage of correct answers (above) and confidence scores for the comprehension quizzes (below).]{
    \resizebox*{7cm}{!}{\includegraphics{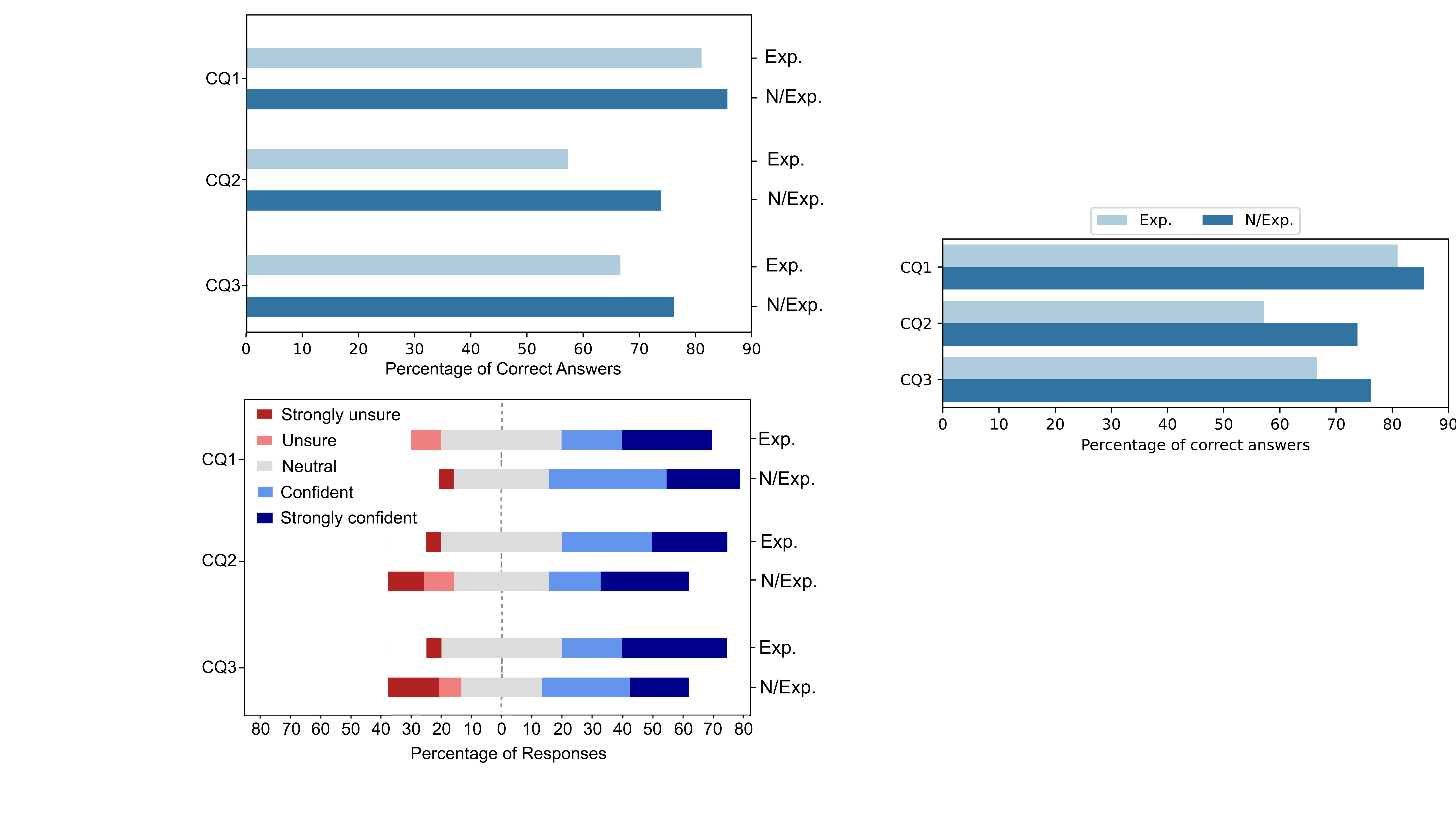}}}
    \caption{Comparison of questionnaire responses between participants with (\textit{Exp.}) and without (\textit{N/Exp})) AI experiences.} 
    \label{fig:comparison_ai_both}
\end{figure}

One objective of this study is to explore whether participants gain technical comprehension through the IML experience. 
Prior knowledge of ML could potentially have a significant impact on this aspect.
Figure~\ref{fig:comparison_ai_both}(a) shows the distribution of usability scores grouped by prior ML knowledge.
The group with no AI experience (\textit{N/Exp.}) tends to rate UQ1, UQ2, and UQ3 lower than the group with AI experience (\textit{Exp.}).
The average scores for \textit{Exp.} and \textit{N/Exp.} are $1.55$ and $1.10$ for UQ1, $1.25$ and $0.73$ for UQ2, $1.35$ and $1.24$ for UQ3, $1.25$ and $1.43$ for UQ4, and $0.95$ and $0.97$ for UQ5.
Although these results suggest that understanding and utilizing the map might pose a technical challenge for inexperienced participants, we also observe that more than half of this group provided positive responses to this item.

\begin{table}
    \tbl{The coding results of CQs between participants with (\textit{Exp.}) and without (\textit{N/Exp})) AI experiences.}
    {\begin{tabular}{ccccc}
        \toprule
                      & Answer & \textit{Exp.} & \textit{N/Exp.} & Total \\
        \midrule
        \multirow{3}{*}{CQ1} & Record sounds or have them memorized & 16 & 36 & 52\\
                           & Updating parameters of the AI model & 1 & 0 & 1\\
                           & Incorrect answer & 4 & 6 & 10\\
        \midrule
        \multirow{3}{*}{CQ2} & Referring the similarities between sounds & 6 & 24 & 30\\
                           & Referring to the particular aspect of sounds & 6 & 7 & 13\\
                           & Incorrect answer & 9 & 11 & 20\\
        \midrule
        \multirow{3}{*}{CQ3} & Increasing the amount of the variety of sounds & 10 & 26 & 36\\
                           & Recording better quality sounds & 4 & 6 & 10\\
                           & Incorrect answer & 7 & 10 & 17\\
        \midrule
        Total & & 21 & 42 & 63 \\
        \bottomrule
    \end{tabular}}
    \label{tab:coding}
\end{table}

The upper portion of Fig.~\ref{fig:comparison_ai_both}(b) shows the percentages of correct answers for each item in the comprehension quiz.
Table~\ref{tab:coding} also shows the distribution of specific answers.
Contrary to our intuition, the \textit{N/Exp.} group overperformed the \textit{Exp.} group on all the items.
One possible explanation could be that their prior knowledge introduced some bias, causing participants in the \textit{Exp.} group to misinterpret the technical aspects of the system.
For instance, in CQ2, some participants in the \textit{Exp.} group suggested that the coordinate axes on the map itself held meaning.
In reality, however, the embedded coordinates are calculated through the t-SNE algorithm based solely on similarities, and the coordinates have no semantic meaning.
Examining the number of correct answers for CQ1, around $86\%$ of the \textit{N/Exp.} group answered correctly. 
For CQ2, \textit{N/Exp.} participants found it easier to consider the similarity of sounds, although fewer answers referred to specific sound features like frequency. 
For CQ3, the \textit{N/Exp.} group was less inclined to improve sound quality and more likely to record additional sounds.

The lower portion of Fig.~\ref{fig:comparison_ai_both}(b) shows the differences in confidence scores. 
The average scores for CQ1, 2, and 3 are $0.70, 0.70, 0.80$ for the \textit{Exp.} group and $0.75, 0.60, 0.26$ for the \textit{N/Exp.} group.
We observe that the \textit{N/Exp.} group tended to assign lower confidence ratings for CQ2 and CQ3. 
This is likely because the answers to these two questions were not explicitly provided during the experience. 
Conversely, the \textit{Exp.} group tended to assign higher confidence ratings, suggesting they believed they understood the key concepts without specific explanations.

\subsection{Differences based on User Behavior}

\begin{figure}[t]
    \centering
    \includegraphics[width=.7\linewidth]{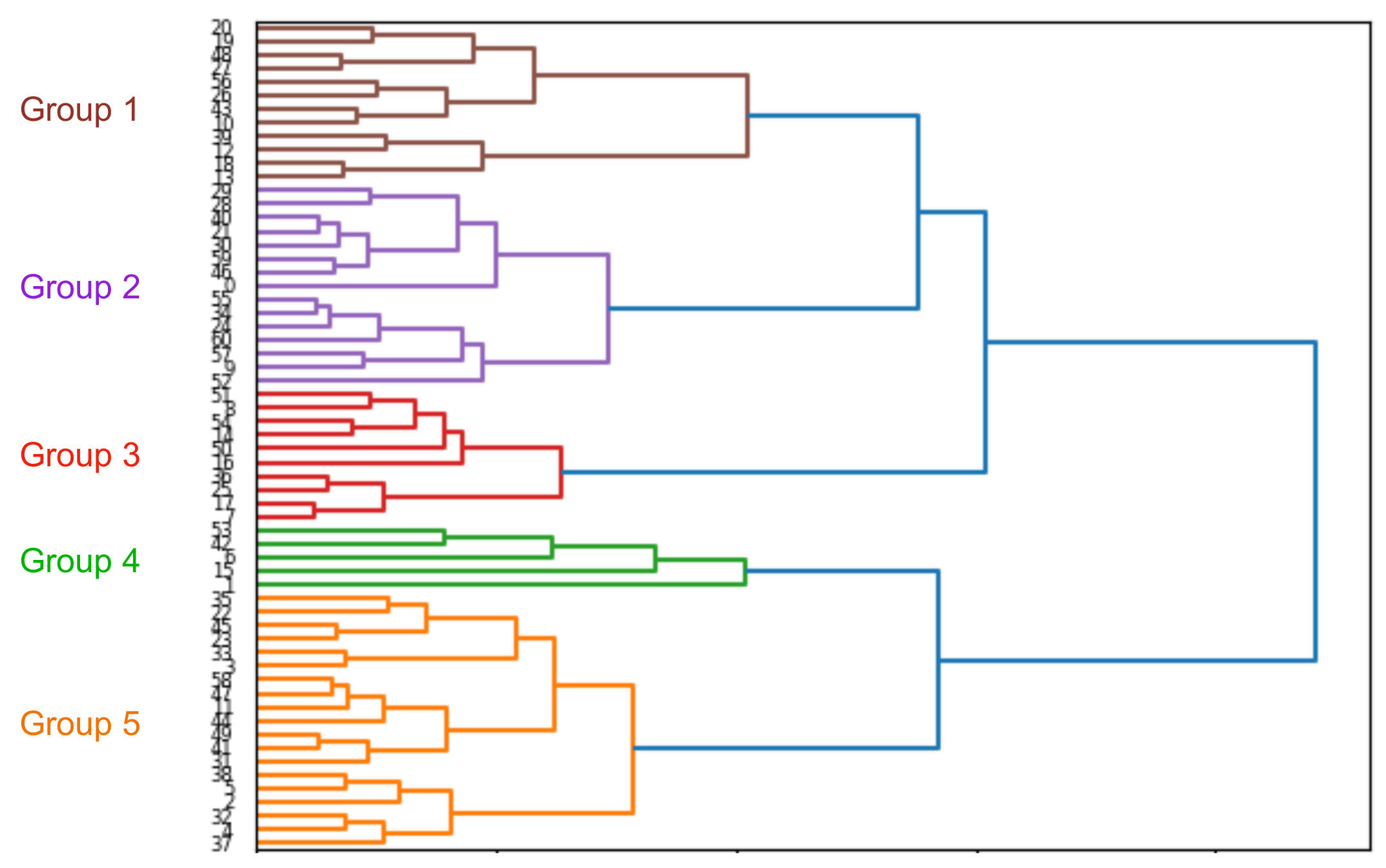}
    \caption{Dendrogram of the participants clustering result based on their interaction logs. Different colors indicate clusters of participants.}
    \label{fig:clustering_tree}
\end{figure}

We conducted further analysis to examine participant differences based on their behavior during the event.
To quantitatively identify differences in system usage from multiple logs, we clustered the participants in a data-driven manner.
We defined 14-dimensional features for each participant based on their interaction logs, with each dimension corresponding to a specific aspect of their interactions.
The primary feature is the number of operations, which includes all log events and their sum: model training and validation, sound recording, and manipulations on the map.
Additionally, the ratio of each item to the total number of clicks is included in the feature set, for example, the ratio of clicks dedicated to recording sounds.
We also incorporated each click count as a measure of time, such as time spent on recording sounds.
Time spent on the map was considered as the time elapsed between a pop-up or sound playback operation and any other operation.
Each feature was then transformed to a $[0.0, 1.0]$ range to maintain consistent minimum and maximum values across participants.
We employed the agglomerative clustering algorithm with the Ward linkage method~\citep{ward1963hierarchical,pedregosa2011scikit}.

\begin{figure}[t]
    \centering
    \subfloat[The ratio of time spent on the feature map with respect to the total time spent on the experience.]{
    \resizebox*{7cm}{!}{\includegraphics{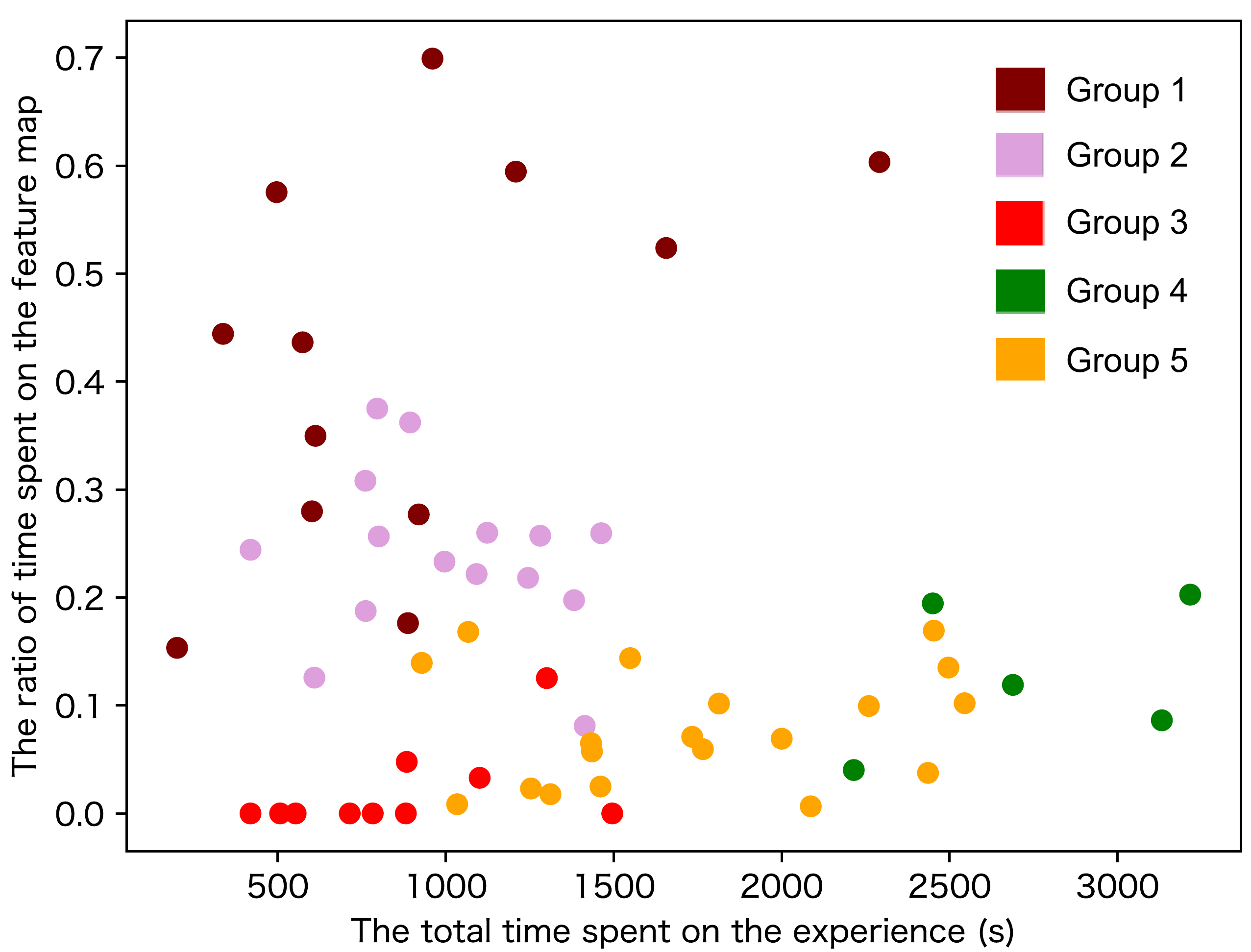}}}
    \subfloat[The time spent on training/validation with respect to the total time spent on the experience.]{\resizebox*{7cm}{!}{\includegraphics{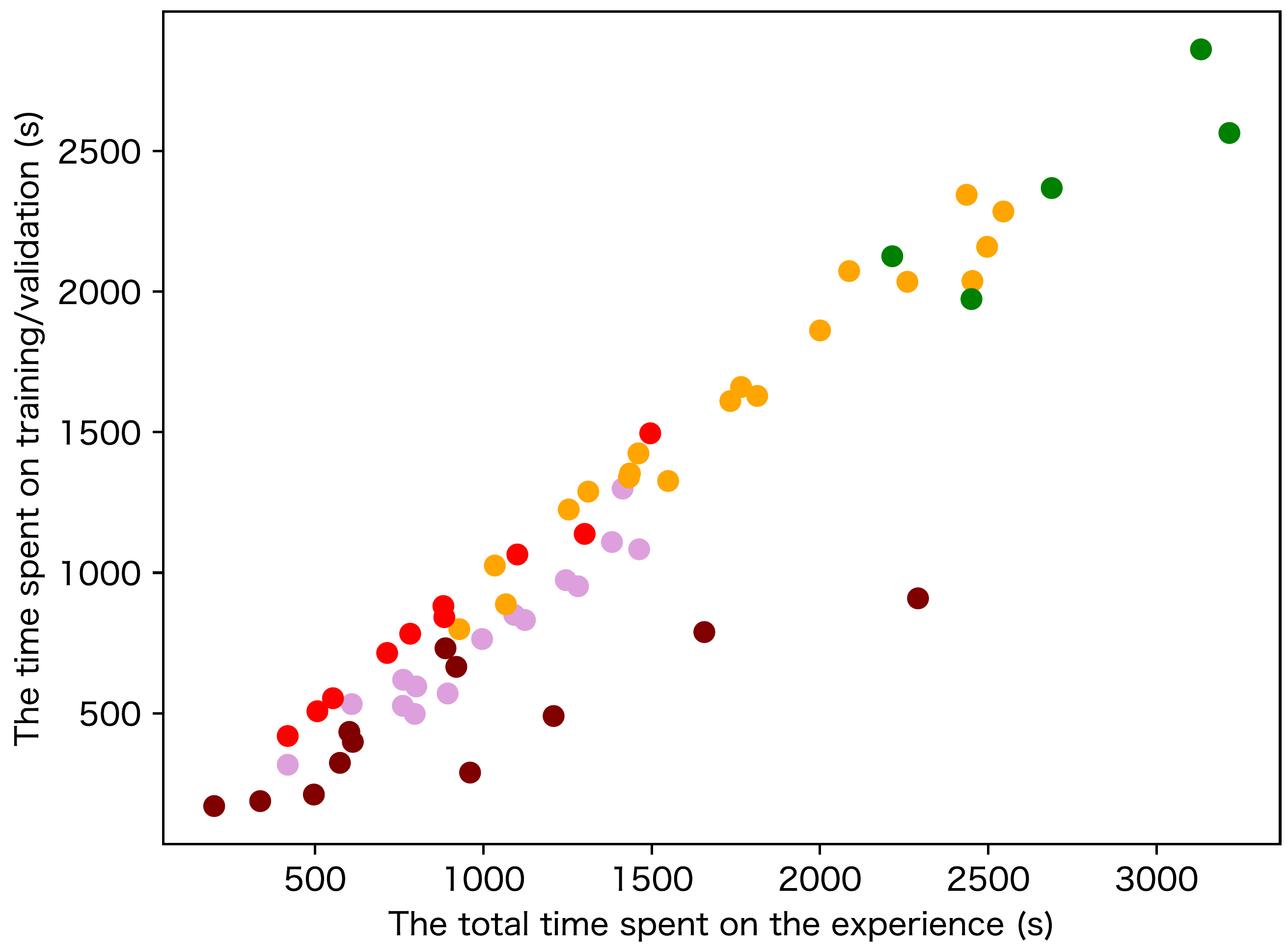}}}
    
    \subfloat[The number of clicks spent on the feature map with respect to the total number of clicks.]{
    \resizebox*{7cm}{!}{\includegraphics{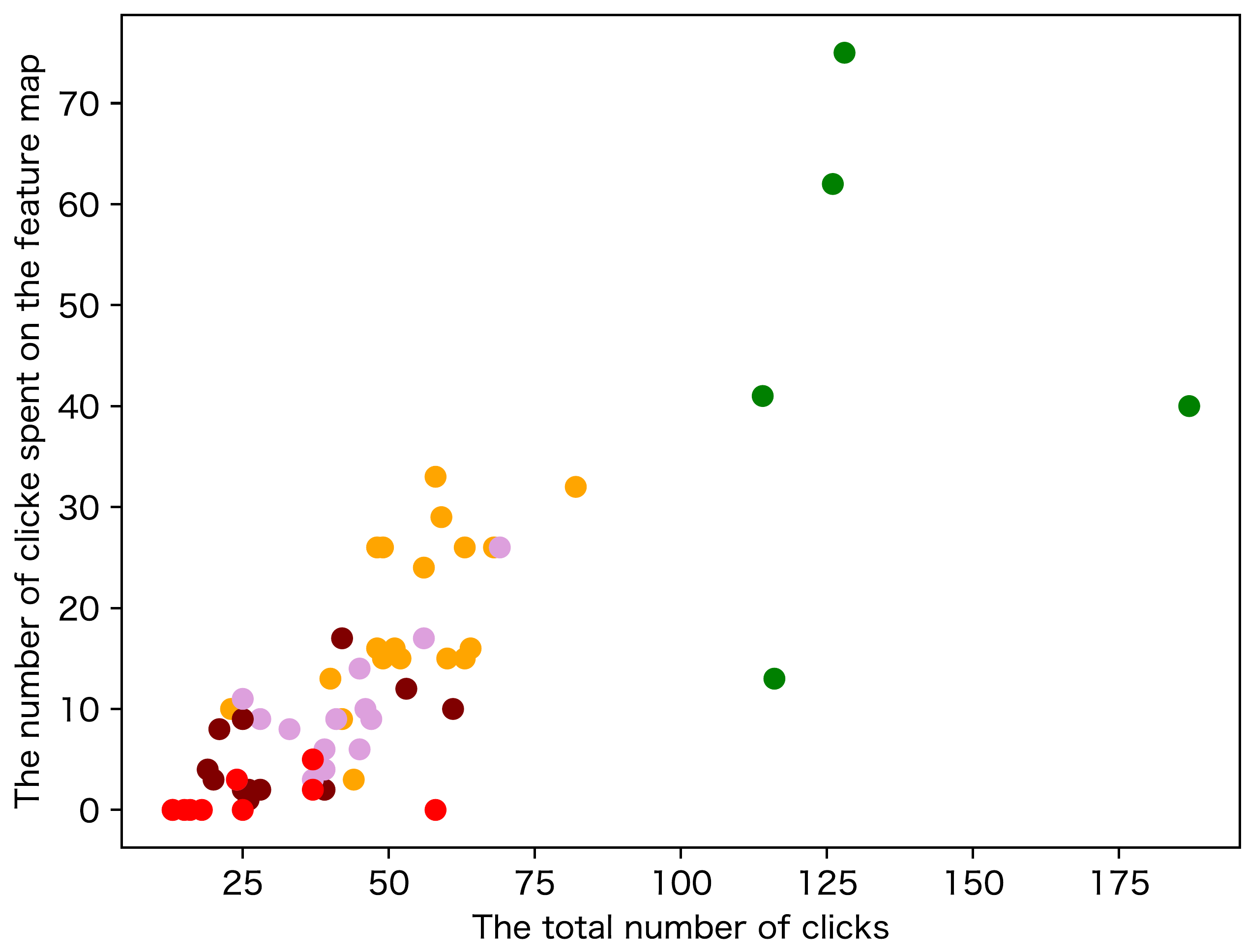}}}
    \subfloat[The ratio of clicks spent on recording sounds with respect to the total number of clicks.]{
    \resizebox*{7cm}{!}{\includegraphics{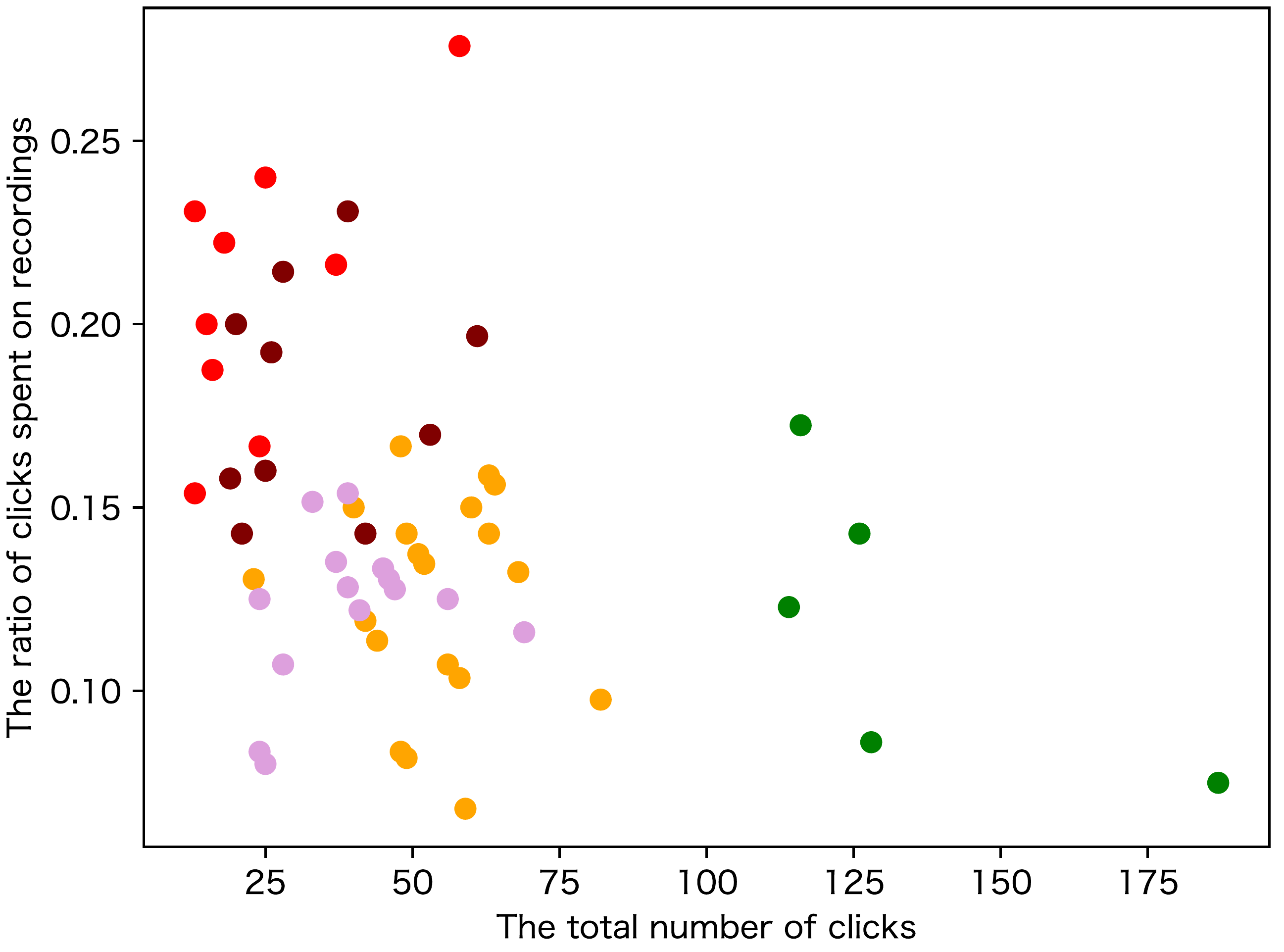}}}
    \caption{Visualization of different usage patterns. Each dot corresponds to one participant, and the color indicates the ID of a cluster shown in Fig.~\ref{fig:clustering_tree}.} 
    \label{fig:clustering_plot}
\end{figure}

\begin{figure}[t]
    \centering
    \subfloat[Scores for the usability questions.]{
    \resizebox*{7cm}{!}{\includegraphics{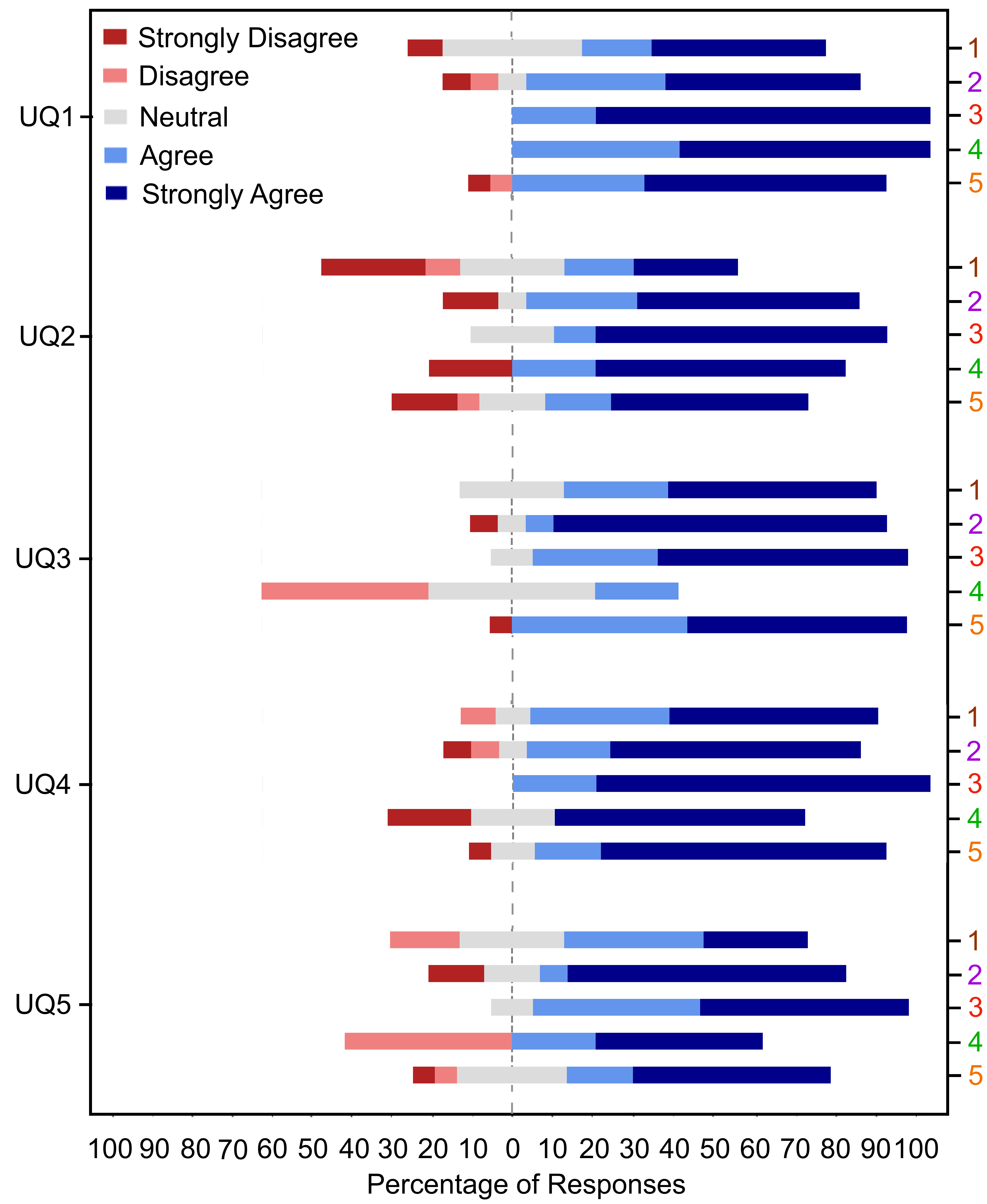}}}
    \subfloat[Percentage of correct answers (above) and confidence scores for the comprehension quizzes (below).]{
    \resizebox*{7cm}{!}{\includegraphics{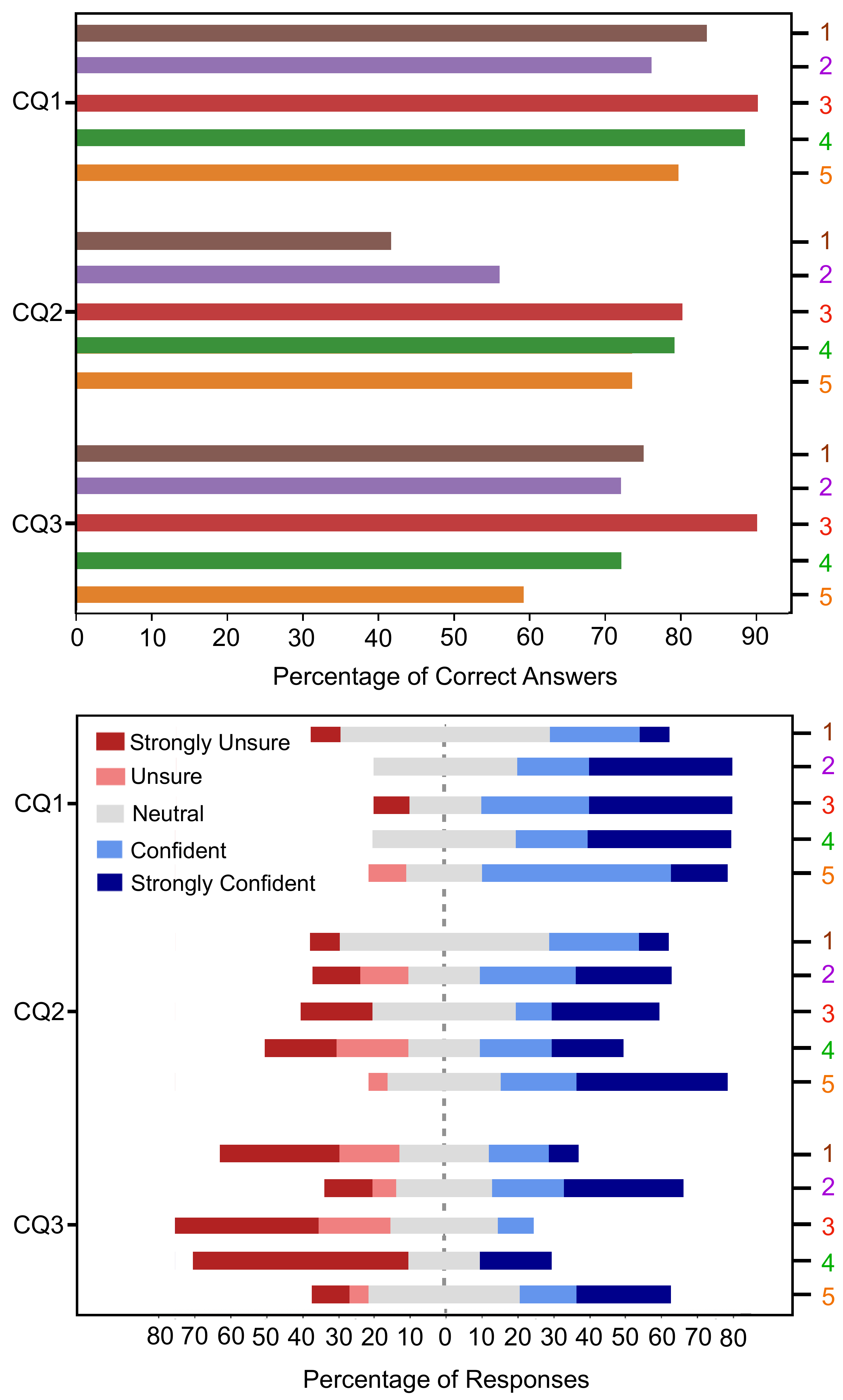}}}
    \caption{Comparison of questionnaire responses between the different clusters. The cluster IDs shown on the right side of the figures correspond to the clustering result in Fig.~\ref{fig:clustering_tree}.} 
    \label{fig:difference_usage}
\end{figure}

Figure~\ref{fig:clustering_tree} shows the dendrogram resulting from the hierarchical clustering.
Each leaf node corresponds to each participant group, and the most similar participants are grouped from bottom to top.
Based on the dendrogram's structure, we identified the five representative clusters, color-coded in Fig.~\ref{fig:clustering_tree}. 
We further applied the Kruskal-Wallis test for these five clusters to determine the most prominent features of these clusters.
Figure~\ref{fig:clustering_plot} presents the visualization of the relationship between the most contributing features.
Among all the features, (1) the ratio of time spent manipulating the feature map, (2) the time spent training and validating the model, (3) the number of clicks related to manipulating the map, (4) the ratio of clicks spent on recording sound out of the total number of clicks, and (5) the total number of clicks contributed the most to cluster determination ($p < 0.0001$).
We visualize (1), (2), (3), and (4) in a way that corresponds to either the total number or the total time.
Since there is a correlation between the total time spent on the experience and (5) the number of clicks ($r = 0.74, p < 0.0001$, Pearson correlation coefficient), we use the total time for (1) and (2) as an alternative to (5).
Figure~\ref{fig:clustering_plot}(a) shows (1), (b) shows (2), (c) shows (3), and (d) shows (4).
Figure~\ref{fig:difference_usage} also shows the differences in questionnaires and quizzes among the behavior clusters.

According to Fig.~\ref{fig:clustering_plot}, group 1 spent a relatively large amount of time on map manipulation and also recorded a comparatively high number of sounds.
This suggests that this group likely observed where the recorded sounds were positioned on the map.
Given that this group scored lower in UQ2 than other groups ($0.08$, $1.06$, $1.50$, $1.00$, and $0.74$ for groups 1, 2, 3, 4, and 5, respectively) according to Fig.~\ref{fig:difference_usage}, group 1 appeared to be more interested in recording and embedding the recorded sounds on the map than investigating the existing samples.
On the other hand, this group struggled to answer CQ2 correctly, which concerns the mechanism of data embedding, compared to the other groups ($42\%$, $56\%$, $80\%$, $79\%$, and $73\%$ for groups 1, 2, 3, 4, and 5, respectively).
Examples of incorrect answers included ``\textit{no idea}'' and those that did not specifically describe the details.
This implies that properly understanding the mechanism of embedding in a map requires deeper insight than merely observing the placements.

Group 2 spent more time on the map, but the number of clicks for recording sounds was lower.
This suggests that group 2 was likely browsing existing samples on the map rather than observing where the recorded sounds were embedded. 
Considering that more than $70\%$ of group 2 scored CQ3 correctly and they had the highest confidence for this quiz ($-0.50$, $0.54$, $-0.90$, $-0.80$, and $0.40$ for groups 1, 2, 3, 4, and 5, respectively), this group may have understood how to improve the model's performance by utilizing the existing samples for training and carefully observing the accuracy.
Note that of the 10 people who made up group 2, four had experience with AI, which is not a large number.
In contrast, group 2 scored the highest in UQ3 ($1.25$, $1.54$, $1.50$, $-0.20$, and $1.36$ for groups 1, 2, 3, 4, and 5, respectively), indicating that they paid attention not only to the samples already on the map but also to their newly recorded samples.

Group 3 spent less time on map manipulation, but the ratio of clicks spent on recordings was relatively higher.
This suggests that group 3 spent more time recording sounds and training the model by using the recorded sounds as training data.
Although the total time and the total number of clicks were lower, this group scored the highest in every CQ, and the scores for usability questions were also higher.
In this sense, this group may be the smartest group that could understand the system specifications and the basic ML concept with less effort.

Group 4, with the longest time and the largest number of clicks, showed distinctive usability and quiz score trends.
They scored the lowest in UQ3 and UQ5, indicating difficulty locating the recorded sounds or creating a desired model.
Note that the scores for UQ5 are $0.65$, $1.12$, $1.40$, $0.60$, and $0.47$ for groups 1, 2, 3, 4, and 5.
They had the lowest confidence in CQ2 ($0.25$, $0.39$, $0.30$, $0.00$, and $0.99$ for groups 1, 2, 3, 4, and 5) and the second-lowest confidence in CQ3.
On the other hand, over $70\%$ of them scored CQ1, 2, and 3 correctly.
Considering these numbers, it is likely that the participants in group 4 underestimate their understanding, but in fact, have taken the most time and number of manipulations to successfully acquire the correct understanding.

Group 5 has basically no specific tendencies for each item, but they tended to wrongly answer CQ3, indicating that their comprehension of how to improve the model's performance was insufficient.
Incorrect answers included the one that focused on the order in which samples are given: ``\textit{let them listen to the easy-to-understand sound first}''.
Others were those that did not specifically describe the details of the idea and those that answered ``\textit{no idea}''.
Given the short duration of the experience relative to the number of clicks, it is possible that they spent more time discussing among members of the group or manipulating YouTube to find sound data.
The short amount of actual time devoted to training/validating the model might have contributed to the lower CQ3 score.

\subsection{Qualitative Observations}

We analyzed the ideas obtained from the worksheet and the interview audio into several groups and observed what trends we found. 
We highlight several noteworthy trends, each of which has at least two or more examples.

\subsubsection*{Creating Detection Models}

Most participants built a simple classifier to distinguish between multiple generic classes. 
However, some participants attempted to create a model specifically designed to detect particular sounds.
A previous study highlighted the challenge non-expert users face in defining a detection task that requires explicit annotation of the background class~\citep{nakao2020use}.
During the interview, one group (P46, group 5) mentioned that they aimed to ``\textit{create an AI model that determines when a railroad crossing is ringing and/or when a train is passing}''.
To address this detection task, they explicitly defined the category of silence:``\textit{we first recorded the sound of crossing, then silence, and finally the sound of trains passing by}''.
The worksheet of another participant group (P58, group 5) revealed their intention to develop an AI model capable of taking photos of fireworks only when cars and people were not obstructing the view.
To achieve this, they trained a model to differentiate the sound of fireworks from other environmental noises.
As stated in their worksheet: ``\textit{we included the sounds of something that is not a firework so that (the AI) would not take pictures when people or vehicles are passing by. It was properly recognized}''.
Some participants in group 5, like these two, focused on creating a model with a unique twist of their own, rather than merely enhancing its performance as a classification model.

\subsubsection*{Interpreting the Feature Map}

Although we only provided a brief explanation of the feature map, some participants attempted to interpret the technical details of the map by themselves.
One group (P12, group 1) wondered why semantic similarities of sound categories were not consistently reflected on the map.
They mentioned that they recorded ``\textit{the sounds of trains and sirens (of ambulances)}'', and discovered that ``\textit{sounds of sirens were close to the sound of musical instruments and sounds of trains were nature sounds}'', even though both are related to vehicles.
While their interpretation is not entirely accurate, they speculated that ``\textit{the distance from the origin (of the coordinate axis of the map)}'' plays a key role in map visualization.
Another group (P44, group 5) also mentioned, ``\textit{we wanted to create a model to classify sounds of different monkeys. We recorded different kinds of monkeys, and then male and female human voices [...] to see if they go to similar locations (on the map)}''.
However, they found these sounds are placed in ``\textit{totally different places.}''
After pondering the reason for this, they speculated that the difference might be due to environmental noise from the videos they used.
They commented, ``\textit{not all the videos are under the same conditions, so if we recorded voices of some monkeys and humans in the same room with the same echoes (they could be embedded to similar locations)}''.

\subsubsection*{Creation of High-Quality Training Data}

We also observed that some participants attempted to improve the quality of their training data.
Recording a specific sound from online videos could result in capturing sounds unrelated to the target sound source.
For example, one participant (P15, group 2) who tried to record the sound of dribbling in basketball mentioned that ``\textit{(the player) took a shot while dribbling, and the sound of the shot hitting the ring was included. [...] I thought it would be better to remove the irrelevant sounds.}''
In response to CQ3, many participants recognized the importance of recording sounds with less environmental noise and fewer task-irrelevant sounds.
Although group 2 generally spent less time training and validating the model, this is just a trend as a whole; for some participants, like P15, their primary interest was to train models with high-quality data.

\label{sec:discussions}

\section{Discussions}
\subsection{Key Findings}
Our hands-on event revealed that there are certain patterns in the participants' level of understanding, ideas, and behavior.
We discuss the findings obtained from the results and the lessons suggested from them. 

\subsubsection{Overall Trends}
According to the questionnaire results, participants generally gained a technical understanding of the system's usage and the underlying principles of ML through the construction of a classification model. 
The high usability scores suggest that the system was user-friendly for participants with varying levels of motivation and prior knowledge. 
Despite the system's novel functionality, the simplicity of required operations for data browsing and model training likely contributed to its high usability.
Participants also scored well on quizzes and demonstrated confidence in their understanding, indicating a basic grasp of ML concepts. 
Although some participants initially struggled to operate the system and explore its functionalities, they successfully resolved these issues through group consultation. 
Notably, many participants engaged in thoughtful considerations directly related to the correct quiz answers through system interaction, even though they were not explicitly provided with quiz content or correct answers during the event.

\subsubsection{Influence of Prior Knowledge on AI}
One of our primary interests was determining whether participants without prior AI/ML experience could understand and engage with the event. 
Although they faced challenges in annotating appropriate samples using the map and improving model performance compared to participants with ML experience, the inexperienced group surprisingly outperformed their experienced counterparts in all comprehension quizzes. 
This outcome suggests that participants without ML experience were able to acquire a certain level of technical comprehension, even without explicit instruction on model design. 
Conversely, participants with ML experience occasionally drew upon past experiences, which led to incorrect answers, highlighting that a fresh perspective might have advantages in understanding new concepts.

A noteworthy observation regarding participants with ML experience concerns their interpretation of the map. 
In the introduction, we only explained that acoustically similar sounds are placed near each other, while dissimilar sounds are positioned further apart. 
Since the sound features are mapped on a two-dimensional feature space, the absolute positions of the sounds hold no meaning. 
However, some participants attributed meaning to the x-axis and y-axis displayed only for understanding relative positions. 
Others assessed similarity based on the semantics of sound sources rather than their acoustic properties. 
Such misunderstandings and misinterpretations can be challenging to address during the event.
Consequently, it is crucial to review the users' understanding post-event and make necessary modifications to ensure a more fruitful interaction and learning experience.

\subsubsection{Differences in System Usage}

When observing system usage, we found that participants exhibited diverse tendencies. 
As this study was conducted as a public event, we noted varying levels of engagement among different groups. 
For instance, some participants frequently recorded sounds and observed their embeddings on the map (group 1), while others focused on understanding the positional relationships of pre-embedded sounds (group 2). 
Additionally, there were differences in the amount of effort invested; some participants completed the experience with minimal manipulation (group 3), while others dedicated considerable time to a trial-and-error process (groups 4 and 5).
These observations highlight the diverse range of user interactions and learning styles exhibited during the event.

By analyzing the differences in behavior concerning questionnaire and quiz responses, trends for each group become more apparent. 
While usability questionnaire scores are generally high, some scores are notably low, such as UQ2 for group 1 and UQ3 and UQ5 for group 4. 
Considering group 1's frequent sound recording, the low UQ2 score may stem from fewer opportunities to use pre-embedded samples on the map. 
In contrast, group 4's extensive system usage suggests that their low UQ3 and UQ5 scores might indicate an underestimation of their own performance or genuine concerns about functional issues.
As usability questionnaires are subjective, we can only speculate on the causes of high or low scores. 
However, if participants used a certain feature extensively and still gave it a low score, it's likely that a problem exists with the functionality. 
For instance, UQ3's low score could be due to the map not rendering smoothly when zooming in and out, while UQ5's low score might be attributed to the difficulty in improving model accuracy with limited data available.

For the comprehension quizzes (CQs), CQ1 was particularly easy to answer, with all groups achieving over $75\%$ correct answers. 
For CQ2, which concerns the position of embedded sounds, groups 3, 4, and 5 had a percentage of correct answers above $70\%$, while groups 1 and 2 scored below $60\%$.
Interaction logs reveal that groups 1 and 2 were especially active on the map, but their understanding of how embeddings are calculated was incorrect.
Since each sample is embedded purely based on the extracted sound features, the embedded location may sometimes appear totally random to participants.
More detailed explanations and visualization would be beneficial to help users interpret embedding results.

For CQ3, only group 5 had a percentage of correct answers below $60\%$, while the others scored above $70\%$. 
Despite CQ3's relatively high percentage of correct answers, all groups exhibited low confidence in answering this question. 
Groups 3 and 4, in particular, showed a high percentage of correct answers but low confidence for CQ3.
Since there are multiple factors involved in training data creation, it is difficult for the participants to be certain whether her/his attempt, e.g., recording more diverse sounds, is really improving the accuracy or not.
This lack of certainty might have led to low confidence scores.
For this part, additional instructions or explanations might be needed as XAI systems offer reasons why the model's accuracy is high or low~\cite{das2020opportunities,adadi2018peeking}.

\subsubsection{Insights into Task Design and Data}
Through qualitative analyses, we observed that some participant groups achieved a high level of technical understanding even without prior ML knowledge. 
The event's location in a science museum might have influenced the participants' background, but this demonstrates the potential for hands-on experiences.
Participants who reached interesting insights were actively engaged in group discussions.
For example, P46, who designed an AI model to detect trains passing by, engaged in deep discussions with family members, one of whom had technical knowledge. 
Similarly, P44, who considered the position of apes and human voices in the embeddings, brainstormed with a friend to verify their hypotheses.
While it is uncertain if group discussions are necessary to reach interesting ideas, exchanging ideas during the event was effective for these participants. 
Another key factor might be that they were likely drawn from an audience with an interest in science, mathematics, and technology. 
Providing opportunities for discussion and collaborative experiences can contribute to enhancing technical understanding.

\subsection{Limitations and Future Work}
One limitation of our study is the difficulty participants faced in correctly understanding the concept of classification through the interactive feature map. 
Although the map visualization aimed to provide insight into the internal mechanism, no direct visualization illustrated how ML-based classification works.
A potential direction for improvement is to incorporate ideas from previous studies on interpretable ML to visualize the classification process, such as those proposed by Selvaraju et al.~\cite{selvaraju2017grad} and Ribeiro et al.~\cite{ribeiro2016should}.
For example, it might be possible to represent classification by drawing a pseudo-boundary line on the map. 
If the boundaries are drawn according to the classification model created by the participants, they can better understand the relationship between the class and the area on the map, such as ``\textit{one side is recognized as the first class, and the other side is recognized as the second class}''. 
This enhanced visualization could facilitate a deeper understanding of the classification process.

Another limitation of this study is the challenge of assessing technical comprehension through free-writing quizzes alone. 
For instance, the ability to provide the correct answer to an open-ended format does not necessarily guarantee the ability to apply that knowledge to one's model design in practice, and vice versa. 
The fact that relatively passive participants scored higher in the comprehension quizzes indicates that quizzes alone might not ensure their technical comprehension.
In future work, various types of comprehensive evaluation metrics should be employed, focusing on how participants' comprehension of ML technology evolves through hands-on experience. 
This could include approaches like coding tests or having participants design their desired model independently using no-code AI development tools. 
By incorporating multiple evaluation methods, a more accurate assessment of participants' technical capabilities can be achieved, providing valuable insights into the effectiveness of such educational experiences.

Finally, extending the study to other ML tasks is an essential future direction. 
Our study focused on sound classification, while other types of task formulations and input modalities would present unique challenges and characteristics. 
It is unclear whether the participants gained an understanding of ML in general from our event, and further technical comprehension could be achieved by allowing them to experience a variety of tasks and inputs.
In particular, the significant development of high-performance language models in recent years~\cite{rombach2022high,rombach2022high,brown2020language} cannot be overlooked, and understanding how people perceive the training and inference process of such models is a crucial topic to explore. 
Some of these models employ prompt engineering~\cite{radford2021learning,liu2023pre}, which allows users to change the model's behavior without additional fine-tuning. 
Prompt instructions represent a fundamentally different paradigm from ML fine-tuning, necessitating the design of interfaces for novice users with a distinct mindset from existing IML systems.
Nonetheless, we believe that the findings of this study remain important when addressing prompt-based models, such as how participants should be facilitated to understand the technology and how events should be designed to achieve this goal.

\label{sec:conclusion}

\section{Conclusion}
In this study, we conducted a public IML hands-on event at a science museum and analyzed how diverse public people gained technical comprehension through such an experience.
Results indicate that a significant portion of the participants successfully acquired a basic understanding of ML, and they were confident in their comprehension, even without prior knowledge of ML.
Analysis of the participants' interaction logs revealed distinct usage patterns, suggesting that not all participants would necessarily use the system as intended.
This study offers valuable insights into the potential of IML for non-expert users and highlights the need for a different design strategy when creating IML systems for such purposes.

\section*{Acknowledgements}
This work was supported by JST CREST Grant Number JPMJCR19F2, Japan. 
We would like to thank Miraikan - the National Museum of Emerging Science and Innovation (especially Bunsuke Kawasaki, Sakiko Tanaka, Koki Sakuma, and Chisa Mitsuhashi) for their great support and cooperation.

\bibliographystyle{apacite}
\bibliography{main}

\end{document}